\documentclass[10pt]{elsarticle}

\usepackage{bm}
\usepackage{mathtools}
\usepackage{graphicx}
\usepackage{multirow}
\newcommand{\norm}[1]{\left\lVert #1 \right\rVert}
\usepackage{algorithm}
\usepackage{algpseudocode}
\usepackage{xcolor}
\usepackage{algpascal}
\usepackage{textcomp}
\usepackage{amsthm}
\theoremstyle{remark}
\newtheorem{rem}{Remark}
\usepackage{tikz}

\usepackage{smartdiagram}
\usepackage[utf8]{inputenc}
\usetikzlibrary{arrows,positioning}
\usepackage{makecell} 
\setcellgapes{5pt}

\usepackage{amsmath,amsfonts,amssymb}

\DeclareMathOperator*{\argmin}{arg\,min}

\usepackage{tikz}

\usepackage{hyperref}

\journal{Fuzzy Sets and Systems}

\begin{document}
	
	\begin{frontmatter}
		
		\title{Fuzzy clustering of circular time series based on a new dependence measure with applications to wind data}


		\author[mymainaddress]{Ángel López-Oriona\corref{mycorrespondingauthor}}
		
		\author[mymainaddress]{Ying Sun}
		\cortext[mycorrespondingauthor]{Corresponding author}
		
		\author[mysecondaryaddress]{Rosa M. Crujeiras}
		
		\address[mymainaddress]{King Abdullah University of Science and Technology (KAUST), Thuwal 23955-6900, Saudi Arabia.}
		\address[mysecondaryaddress]{Department of Statistics, Mathematical Analysis and Optimization, Universidade de Santiago de Compostela, Santiago de Compostela 15782, Spain}

		
		\begin{abstract}
			Time series clustering is an essential machine learning task with applications in many disciplines. While the majority of the methods focus on time series taking values on the real line, very few works consider time series defined on the unit circle, although the latter objects frequently arise in many applications. In this paper, the problem of clustering circular time series is addressed. To this aim, a distance between circular series is introduced and used to construct a clustering procedure. The metric relies on a new measure of serial dependence considering circular arcs, thus taking advantage of the directional character inherent to the series range. Since the dynamics of the series may vary over the time, we adopt a fuzzy approach, which enables the procedure to locate each series into several clusters with different membership degrees. The resulting clustering algorithm is able to group series generated from similar stochastic processes, reaching accurate results with series coming from a broad variety of models. An extensive simulation study shows that the proposed method outperforms several alternative techniques, besides being computationally efficient. Two interesting applications involving time series of wind direction in Saudi Arabia highlight the potential of the proposed approach.
		\end{abstract}
		
		\begin{keyword}
			Circular time series; Clustering; Fuzzy logic; Fuzzy $C$-medoids; Wind direction
		\end{keyword}

		
	\end{frontmatter}

	\section{Introduction}\label{sectionintroduction}
	
	Time series clustering refers to the problem of splitting a set of unlabelled time series into homogeneous groups in such a way that similar series are placed together in the same group and dissimilar series are located in different groups. Usually, the clustering task is driven by the desired similarity notion, which can be established in very different manners. Frequently, the purpose is to identify groups associated with similar generating models, thus allowing to characterize a few dynamic patterns without the need to analyze and model each single time series in the database, which is rarely the objective when dealing with a large number of series. The complexity inherent to objects evolving over time, such as determining a suitable dissimilarity principle, existence of regime shifts and dealing with series of unequal length, together with the vast range of applications where time series clustering plays an essential role, justify a growing interest in this challenging topic. Comprehensive overviews including recent advances, future prospects, classical references and specific application areas are provided by \cite{liao2005clustering, aghabozorgi2015time, maharaj2019time}.
	
	The majority of clustering methods focus on real-valued time series. Among them, there are techniques based on discriminating between different geometric profiles in the time series \cite{izakian2015fuzzy, luczak2016hierarchical}, model-based approaches \cite{corduas2008time, d2013autoregressive, d2016garch}, feature-based procedures \cite{d2009autocorrelation, d2012wavelets, lopez2022quantile1, lopez2022quantile2}, or techniques based on reducing the dimensionality of the original time series as a preliminary step \cite{singhal2005clustering, pealat2021improved}. The suitability of each class of algorithms usually depends on the nature of the time series and the final goal of the user, with no approach dominating the remaining ones in every possible context. According to the cluster assignment criterion, two different paradigms are considered according to whether a “hard” or “soft” partition is constructed. Traditional clustering leads to hard solutions, where each data object is located in exactly one cluster. Overlapping groups are not allowed by hard partitions, which may not be suitable for many real life applications where the cluster boundaries are not clearly determined or some objects are equidistant from some clusters. Soft clustering techniques provide a more versatile approach by considering the concept of membership degrees. In a soft partition, every object has an associated vector of membership degrees indicating the amount of confidence in the assignment to each of the respective clusters. A well-known approach to perform soft clustering is via fuzzy clustering methods \cite{miyamoto2008algorithms}. Adoption of the fuzzy approach is usually advantageous when analyzing time series datasets, since regime shifts are frequent in practice. This becomes a challenge when dealing with a large number of time series due to the high level of complexity of the underlying fuzzy partition. 
	
	A considerably smaller number of works have dealt with the task of clustering time series having a range different from the real-line. Among them, we can highlight some clustering algorithms for count time series \cite{etienne2014model, cerqueti2022ingarch} or categorical time series \cite{pamminger2010model, garcia2015framework}, which are frequently employed for analyzing biological sequences. However, to the best of our knowledge, there does not exist not a single work in the literature addressing the problem of clustering circular time series (CTS), i.e., time series having the unit circle as range \cite{fisher1994time, holzmann2006hidden}, which is surprising, since these objects appear quite frequently in several fields. Specifically, one of the most common types of CTS are time series of wind direction \cite{breckling2012analysis, harvey2023modelling}, which are usually considered in environmental science to analyze several important phenomena as climate change or the characterization of ocean surface wind \cite{plecha2023offshore}, among many others. 
	
	Previous considerations clearly highlight the need for clustering algorithms specifically designed to deal with CTS. It is worth highlighting that CTS clustering involves several challenges, including the inability of standard statistical methods to analyze circular data \cite{jammalamadaka1988correlation}, the difficulty inherent to the definition of dependence measures in a circular context, and the lack of works on the topic which are available in the literature, among others. In addition, given the complex nature of time series databases, it is beneficial to develop the fuzzy approach, whose versatility allows to capture changes in the dynamic behaviors of the series over time in real applications. 
	
	The main objective of this work is to introduce fuzzy clustering algorithms for CTS being able to: (i) group together circular series coming from similar underlying processes, (ii) reach accurate results with series generated from a wide variety of circular processes, and (iii) efficiently perform the clustering task from a computational perspective. To this aim, we first introduce a distance between CTS. Since our goal is to group the time series according to their underlying structures, the dissimilarity relies on extracted features providing information about the serial dependence patterns. The computed features are based on a new measure of serial dependence which considers circular arcs of a given center and radius, and, thus, the metric takes advantage of the directional character inherent to the series range. The dissimilarity is used as input to the standard fuzzy $C$-medoids algorithm, which allows for the assignment of gradual memberships of the CTS to clusters. Assessment of the clustering approach is carried out through an extensive simulation study by generating data from different classes of circular processes. The performance of some alternative dissimilarities designed to deal with circular or real-valued time series is also examined for comparison purposes. 
	
	The usefulness of the proposed approach is illustrated by means of two case studies involving a dataset of hourly time series of wind direction in several locations of Saudi Arabia between the years 2010 and 2017. In both cases, we construct the individual time series in a way that each one of them includes one whole month of data. In the first application, we consider the summer months as June, July, August and September, and the winter months as December, January, February and March, for a specific location. The resulting clustering partition indicates that the proposed algorithm is able to accurately differentiate between winter and summer time series. Moreover, there are some series showing a considerable fuzzy behavior, which means that they share the patterns of both seasons almost in equal measure, and could deserve a careful analysis. For the second application, we consider time series in two different locations, and show that the proposed approach is also able to detect the corresponding geographical differences to some extent. In sum, the considered case studies show that: (i) the new measure of serial dependence can provide a significant understanding about the nature of the time series under consideration, and (ii) the clustering algorithm can produce a meaningful partition whose membership degrees can provide insights to practitioners about certain time series in the dataset. 
	
	The rest of the paper is organized as follows. A new measure of serial dependence for circular processes and a distance between CTS based on a proper estimate of this measure are introduced in Section \ref{sectiondistance}, where a simple example is also shown to illustrate the usefulness of the dissimilarity. In Section \ref{sectionfuzzyalgorithm}, a fuzzy clustering algorithm based on the proposed metric is presented. The performance of the technique is analyzed in Section \ref{sectionsimulations} through a comprehensive simulation study where several alternative procedures are also assessed. Section \ref{sectionapplications} presents the application of the proposed technique to the wind direction dataset. Some concluding remarks are summarized in Section \ref{sectionconclusions}.\footnote{\textcolor{black}{The code used for the experiments in this paper is available in \href{https://github.com/anloor7/PostDoc/tree/main/r_code/circular}{github/paper\_circular}}.}

	\section{A new distance between circular time series}\label{sectiondistance}
	
	In this section, a distance between circular series is defined after introducing a new measure of serial dependence for this kind of series. The advantages of the proposed distance with respect to metrics based on classical correlation measures for circular data are properly motivated through an interesting example.  
	
	\subsection{Measuring serial dependence in circular time series}\label{subsectionmeasuring}
	
	Let $\{\Theta_t, t \in \mathbb{Z}\}$ a strictly stationary circular stochastic process (i.e., each $\Theta_t$ is a circular random variable) whose variables are measured as angles with reference to the same zero direction and in the clockwise rotation. Let us assume that each variable $\Theta_t$ has a unique circular median $\tilde{\mu} \in [0, 2\pi)$ (see Section 3.2.2 in \cite{fisher1995statistical}) and denote by $q_p$ the corresponding circular quantile of level $p\in [0, 1]$. Specifically, given a circular random variable $\Theta$ with density $f$, $q_p$ is defined by
	
	\begin{equation}\label{cquantile}
		\int_{\tilde{\mu}-\pi}^{q_p}f(\theta)d\theta=p. 
	\end{equation}

	For a fixed lag, $l \in \mathbb{Z}$, and a pair of probability levels, $\left(\tau, \tau^{\prime}\right) \in[0,1]^2$, consider the covariance of the indicator functions $I\left(\Theta_t \in A_{\tau, r}\right)$ and $I\left(\Theta_{t+l} \in A_{\tau', r}\right)$ given by 
	
	\begin{equation}
		\begin{split}
			\gamma(\tau, \tau', l, r)=\operatorname{Cov}\Big[I\left(\Theta_t\in A_{\tau, r}\right), I\left(\Theta_{t+l} \in A_{\tau', r}\right)\Big] \\ =P(\Theta_t\in A_{\tau, r}, \Theta_{t+l}\in A_{\tau', r})-P(\Theta_t\in A_{\tau, r})P(\Theta_{t+l}\in A_{\tau', r}),
		\end{split}
	\end{equation}

	\noindent where $A_{p, w}$ denotes the arc of center $q_p$ and radius $w \in [0, \pi)$ on the unit circle, that is,
	
	\begin{equation}
		\begin{split}
			A_{p, w}=\begin{cases}
				A_{p, w}^* & \text{if \, \,}   (q_p-w)>0 \text{\, \, and \, \,} (q_p + w)< 2\pi, \\
				\overline{A}_{p, w}^* & \text{otherwise},
			\end{cases}
		\end{split}
	\end{equation}
	
	\noindent where $\overline{(\cdot)}$ denotes the complimentary set operator on the unit circle and
	
	\begin{equation}
		A_{p, w}^*=\big\{\Psi \in [0, 2\pi): \Psi_1(p, w) \leq \Psi \leq \Psi_2(p, w)\big\},
	\end{equation}
	
	\noindent with
	
	\begin{align}
		\begin{split}
			\Psi_1(p, w)& = \min\{(q_p-w)\mod(2\pi), (q_p+w)\mod(2\pi)\}, \\ \Psi_2(p, w) & = \max\{(q_p-w)\mod(2\pi), (q_p+w)\mod(2\pi)\}.
		\end{split}
	\end{align}
	
	As the quantity $\gamma(\tau, \tau', l, r)$ can be seen as an extension of the classical quantile autocovariance to the circular case, we call this quantity the \textit{circular quantile autocovariance of lag $l$ and radius $r$ for levels $\tau$ and $\tau'$}. Note that $\gamma(\tau, \tau', l, r)$ can be normalized to the interval $[-1, 1]$ by considering the quantity 
	
	\begin{align}
		\begin{split}
			& \rho(\tau, \tau', l, r)  = \\  & \frac{\gamma(\tau, \tau', l, r)}{\Big[P(\Theta_t\in A_{\tau, r})P(\Theta_{t+l}\in A_{\tau', r})\big[1-P(\Theta_t\in A_{\tau, r})\big]\big[1-P(\Theta_{t+l}\in A_{\tau', r})\big]\Big]^{1/2}}, 
		\end{split}
	\end{align}
	
	\noindent which we refer to as \textit{circular quantile autocorrelation} (CQA) \textit{of lag $l$ and radius $r$ for levels $\tau$ and $\tau'$}. 
	
	In practice, quantities $\gamma(\tau, \tau', l, r)$ and $\rho(\tau, \tau', l, r)$ must be estimated from a $T$-length realization of the circular process $\{\Theta_t, t \in \mathbb{Z}\}$, $\boldsymbol \Theta_T=(\theta_1, \theta_2, \ldots, \theta_T)$, often referred to as \textit{circular time series} (CTS). Estimates of these quantities can be obtained by considering natural estimates of the probabilities $P(\Theta_t\in A_{p, w})$ and $P(\Theta_t\in A_{p, w}, \Theta_{t+l}\in A_{p', w})$ given by
	
	\begin{align}
		\begin{split}
			&	\widehat{P}(\Theta_t\in A_{p, w})  =\frac{1}{T}\sum_{i=i}^{T}I(\theta_i\in \widehat{A}_{p, w}), \\ & \widehat{P}(\Theta_t\in A_{p, w},  \Theta_{t+l}\in A_{p', w})  =\frac{1}{T-l}\sum_{i=1}^{T-l}I(\theta_i\in \widehat{A}_{p, w})I(\theta_{i+l}\in \widehat{A}_{p', w}),
		\end{split}
	\end{align}
	
	\noindent respectively, where $\widehat{A}_{p, w}$ is the arc of center $\widehat{q}_p$ and radius $w$, being $\widehat{q}_p$ the standard estimate of the $p$th quantile of a circular random variable as defined in \eqref{cquantile}. The corresponding estimates for $\gamma(\tau, \tau', l, r)$ and $\rho(\tau, \tau', l, r)$ are denoted as $\widehat{\gamma}(\tau, \tau', l, r)$ and $\widehat{\rho}(\tau, \tau', l, r)$, respectively.

	\subsection{A novel dissimilarity between circular time series}\label{subsectiondissimilarity}
	
	Suppose we have two stationary circular processes $\{\Theta_t^{(1)}, t \in \mathbb{Z}\}$ and $\{\Theta_t^{(2)}, t \in \mathbb{Z}\}$. A simple dissimilarity criterion to measure discrepancy between both processes consists in comparing their representations in terms of CQA of a fixed radius $r$ evaluated for several lags on a common set of selected probability levels. Specifically, given a collection of $L$ lags, $\mathcal{L}=\{l_1, l_2, \ldots, l_L\}$, and a collection of $P$ probability levels, $\mathcal{T}=\{\tau_1, \tau_2, \ldots, \tau_P\}$, we define a distance $d_{CQA}$ as 
	
	\begin{equation}\label{dcqa}
		d_{CQA}\big(\Theta_t^{(1)}, \Theta_t^{(2)}\big)=\frac{1}{4LP^2}\sum_{k=1}^{L}\sum_{i=1}^{P}\sum_{j=1}^{P}\Big(\rho(\tau_i, \tau_j, l_k, r)^{(1)}-\rho(\tau_i, \tau_j, l_k, r)^{(2)}\Big)^2,
	\end{equation}

	\noindent where the superscripts (1) and (2) indicate that the corresponding quantities refer to the processes $\{\Theta_t^{(1)}, t \in \mathbb{Z}\}$ and $\{\Theta_t^{(2)}, t \in \mathbb{Z}\}$, respectively. Note that $d_{CQA}\big(\Theta_t^{(1)}, \Theta_t^{(2)}\big) \in [0,1]$. 
	
	In practice, $d_{CQA}$ must be estimated on the basis of realizations $\boldsymbol \Theta_{T_1}^{(1)}=(\theta_1^{(1)}, \theta_2^{(1)}, \ldots, \theta_{T_1}^{(1)})$ and $\boldsymbol  \Theta_{T_2}^{(2)}=(\theta_1^{(2)}, \theta_2^{(2)}, \ldots, \theta_{T_2}^{(2)})$ of the processes $\{\Theta_t^{(1)}, t \in \mathbb{Z}\}$ and $\{\Theta_t^{(2)}, t \in \mathbb{Z}\}$, respectively, with corresponding lengths $T_1$ and $T_2$, by means of 
	
	\begin{equation}\label{estimateddcqa}
		\widehat{d}_{CQA}\big(\Theta_t^{(1)}, \Theta_t^{(2)}\big)= \frac{1}{4LP^2}\sum_{k=1}^{L}\sum_{i=1}^{P}\sum_{j=1}^{P}\Big(\widehat{\rho}(\tau_i, \tau_j, l_k, r)^{(1)}-\widehat{\rho}(\tau_i, \tau_j, l_k, r)^{(2)}\Big)^2,
	\end{equation}
	
	\noindent where the superscripts (1) and (2) indicate that the corresponding estimates are computed from the realizations $\boldsymbol \Theta_{T_1}^{(1)}$ and $\boldsymbol  \Theta_{T_2}^{(2)}$, respectively. 
	
	Three remarks concerning the proposed dissimilarity are provided below. 
	
	\begin{rem}
		\label{remquantiles}
		\textit{On the choice of the quantiles}. Note that the distance $d_{CQA}$ and its estimate are based on the quantiles in \eqref{cquantile}, which are inherent to the definition of the quantity $\rho(\tau, \tau', l, r)$. In this regard, alternative definitions of quantiles (e.g., quantiles based on depth measures for circular data \cite{liu1992ordering, buttarazzi2018boxplot}) could be used to define $\rho(\tau, \tau', l, r)$, and, in turn, $d_{CQA}$ and its estimate. In fact, all the numerical experiments shown throughout the paper (see Section \ref{sectionsimulations}) were also carried out by considering several counterparts of $\widehat{d}_{CQA}$ associated with different definitions of quantiles. Overall, the clustering accuracy attained with these alternative distances was not significantly different than from the one achieved with $\widehat{d}_{CQA}$. For this reason, we decided to define the quantity $\rho(\tau, \tau', l, r)$ using the quantiles in \eqref{cquantile}. 
	\end{rem}
	
	\begin{rem}
		\label{remfeaturemetric}
		\textit{Advantages of feature-based distances}. The metric $\widehat{d}_{CQA}$ belongs to the class of feature-based distances, since it is based on comparing extracted features. The discriminative capability of this kind of distances mostly depends on selecting suitable features for a given context. If a proper set of features is used, then these distances present very nice properties such as dimensionality reduction, low computational complexity, robustness to the generating models, and versatility to compare series with different lengths. It is worth remarking that these properties are not satisfied by other dissimilarities between time series. For instance, metrics based on raw data usually involve high computational cost and require the series to have the same length, while model-based dissimilarities are expected to be extremely sensitive to model misspecification.
	\end{rem}
	
	\begin{rem}
		\label{remhyperparametersd}
		\textit{Hyperparameters in $\widehat{d}_{CQA}$}. Computation of $\widehat{d}_{CQA}$ requires the selection of three hyperparameters, namely the set of lags, $\mathcal{L}$, the collection of probability levels, $\mathcal{T}$, and the radius, $r$. As the distance $\widehat{d}_{CQA}$ is used in this work to perform fuzzy clustering of CTS, we propose to select these parameters in a way that the clustering accuracy gets maximized for a given set of time series. The specific hyperparameter selection procedure is described in Remark \ref{remhyperparameters} of Section \ref{subsectionfuzzyalgorithm}.  
	\end{rem}
	
	\subsection{Motivating example}\label{subsectionmotivatingexample}
	
	This section illustrates the advantages of using the distance $\widehat{d}_{CQA}$ to differentiate between circular time series. Let us consider two real-valued processes $\{X_t^{(1)}, t \in \mathbb{Z}\}$ and $\{X_t^{(2)}, t \in \mathbb{Z}\}$ defined by the recursions

	\begin{align}\label{qar}
		\begin{split}
			&	X_t^{(1)}=0.2(U_t-0.5)X_{t-1}^{(1)}+1.2(U_t-0.5)X_{t-2}^{(1)}+\Phi^{-1}(U_t), \\
			&	X_t^{(2)}=-0.2(V_t-0.5)X_{t-1}^{(2)}-1.2(V_t-0.5)X_{t-2}^{(2)}+\Phi^{-1}(V_t),
		\end{split}
	\end{align}  
	
	\noindent where both $\{U_t,  t \in \mathbb{Z}\}$ and $\{V_t,  t \in \mathbb{Z}\}$ are sequences of i.i.d. standard uniform random variables and $\Phi^{-1}$ denotes the quantile function of a standard normal random variable. The previous processes belong to the class of quantile autoregressive (QAR) models proposed by \cite{koenker2006quantile}, which are able to capture systematic influences of conditioning variables on the location, scale, and shape of the conditional distribution of the response. Note that, in the previous processes, $X_t^{(i)}$ and $X_{t-j}^{(i)}$ are uncorrelated, $i,j=1,2$. 
	
	Let us now transform the processes in \eqref{qar} into circular stochastic processes. To this aim, we employ the modulus operator (see Section 2.2 in \cite{fisher1994time}), obtaining the so-called wrapped processes $\{\Theta_t^{(1)}, t \in \mathbb{Z}\}$ and $\{\Theta_t^{(2)}, t \in \mathbb{Z}\}$, with $\Theta_t^{(i)}=X_t^{(i)}\mod(2\pi)$, $i=1,2$. In order to measure dissimilarity between realizations of the previous processes, we will employ the distance $\widehat{d}_{CQA}$ in \eqref{estimateddcqa} and two dissimilarities based on classical correlation coefficients for circular data adapted to the temporal context. The first metric is based on the correlation coefficient of \cite{fisher1983correlation}, whose natural estimate given a realization $\boldsymbol \Theta_T=(\theta_1, \theta_2, \ldots, \theta_T)$ and a lag $l \in \mathbb{Z}$ takes the following form in the temporal setting:
	
	\begin{equation}
		\widehat{\rho}_{FL}(l)=\frac{\sum_{i=1}^{T-l}\sum_{\substack{j=1: \\ j>i}}^{T-l}\sin(\theta_i-\theta_j)\sin(\theta_{i+l}-\theta_{j+l})}{\Big[\sum_{i=1}^{T-l}\sum_{\substack{j=1: \\ j>i}}^{T-l}\sin^2(\theta_i-\theta_j)\Big]^{1/2}\Big[\sum_{i=1}^{T-l}\sum_{\substack{j=1: \\ j>i}}^{T-l}\sin^2(\theta_{i+l}-\theta_{j+l})\Big]^{1/2}}.
	\end{equation}
	
	The second dissimilarity relies on the correlation coefficient of \cite{jammalamadaka1988correlation}. Given $l \in \mathbb{Z}$, a reasonable estimate of this coefficient in the temporal context is given by
	
	\begin{equation}
		\widehat{\rho}_{JS}(l)=\frac{\sum_{i=1}^{T-l}\sin(\theta_i-\overline{\theta})\sin(\theta_{i+l}-\overline{\theta})}{\Big[\sum_{i=1}^{T-l}\sin^2(\theta_i-\overline{\theta})\Big]^{1/2}\Big[\sum_{i=1}^{T-l}\sin^2(\theta_{i+l}-\overline{\theta})\Big]^{1/2}},
	\end{equation}
	
	\noindent where $\overline{\theta}$ is the estimated circular mean (see \cite{mardia2000directional}) considering the realization $\boldsymbol{\Theta}_T$. 
	
	The estimated distances associated with the coefficients $\widehat{\rho}_{FL}(l)$ and $\widehat{\rho}_{JS}(l)$ are defined as
	
	\begin{align}\label{distancecorrelations}
		\begin{split}
			&	\widehat{d}_{FL}\big(\Theta_t^{(1)}, \Theta_t^{(2)}\big)=\frac{1}{4L}\sum_{i=1}^{L}\Big(\widehat{\rho}_{FL}(l_k)^{(1)}-\widehat{\rho}_{FL}(l_k)^{(2)}\Big)^2, \\
			&	\widehat{d}_{JS}\big(\Theta_t^{(1)}, \Theta_t^{(2)}\big)=\frac{1}{4L}\sum_{i=1}^{L}\Big(\widehat{\rho}_{JS}(l_k)^{(1)}-\widehat{\rho}_{JS}(l_k)^{(2)}\Big)^2, 
		\end{split}
	\end{align}  
	
	\noindent respectively, for a fixed set of lags $\mathcal{L}=\{l_1, l_2, \ldots, l_L\}$. 
	
	In order to compute the distance between the wrapped processes $\{\Theta_t^{(1)}, t \in \mathbb{Z}\}$ and $\{\Theta_t^{(2)}, t \in \mathbb{Z}\}$, we simulated 1000 pairs of realizations of length $T=5000$, and calculated the value of the estimated distances $\widehat{d}_{CQA}$, $\widehat{d}_{FL}$ and $\widehat{d}_{JS}$. In all cases, we considered the set of lags $\mathcal{L}=\{1, 2\}$, since these are the lags defining both processes in \eqref{qar}. The set of probability levels $\mathcal{T}=\{0.1, 0.5, 0.9\}$ and a grid of values for $r \in [0, \pi)$, namely $r \in \{0.1, 0.2, \ldots ,3.1\}$, was employed for the computation of $\widehat{d}_{CQA}$. Since the computed values were rather small, the sample elements were multiplied by 100 for the sake of illustration. Table \ref{tablequantities} provides the results of this experimental study. Specifically, the sample mean, standard deviation, and the 5th and 95th quantiles were obtained for each sample of distances. The results associated with the metric $\widehat{d}_{CAQ}$ correspond to the radius $r$ which gave rise to the maximum value of the sample mean, namely $r=1.8$.

	\begin{table}
		\centering 
		\begin{tabular}{cccccc}\hline 
			Distance & Mean & Standard deviation & 5th quantile & 95th quantile &  \\ \hline  
			$\widehat{d}_{CAQ}$	&   0.8922   &   0.0448    &   0.8172    &     0.9652          &  \\
			$\widehat{d}_{FL}$	&   0.0038   &    0.0038     &  0.0002   &      0.0111         &  \\
			$\widehat{d}_{JS}$	&   0.0132   &   0.0106       &  0.0010  &       0.0339        & \\ \hline 
		\end{tabular}
		\caption{Sample mean, standard deviation and quantiles (scaled by a factor of 100) based on a sample of size 1000 for three distances.}
		\label{tablequantities}
	\end{table} 
	
	According to Table \ref{tablequantities}, the correlation-based dissimilarities $\widehat{d}_{FL}$ and $\widehat{d}_{JS}$ are not able at all to differentiate between the corresponding circular time series. This is because the uncorrelatedness existing in the real-valued processes $\{X_t^{(1)}, t \in \mathbb{Z}\}$ and $\{X_t^{(2)}, t \in \mathbb{Z}\}$ is inherited by the wrapped processes $\{\Theta_t^{(1)}, t \in \mathbb{Z}\}$ and $\{\Theta_t^{(2)}, t \in \mathbb{Z}\}$. Therefore, these metrics are not capable of distinguishing between time series when the differences are provoked by forms of nonlinear circular dependence. On the contrary, as the distance $\widehat{d}_{CAQ}$ assesses the possible dependence considering several pairs of arcs on the unit circle, this metric can detect the existing differences in the dynamic behaviors of both time series. 
	
	\begin{figure}
		\centering
		\includegraphics[width=0.8\textwidth]{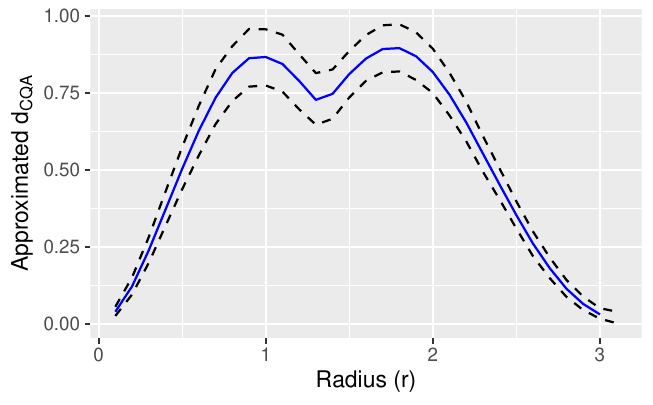}
		\caption{Mean of 1000 independent replicates of $\widehat{d}_{CQA}$, scaled by a factor of 100 (blue curve), as a function of the radius $r$. The lower (upper) dashed line represents the 5th (95th) quantile of the 1000 replicates.}
		\label{curvesdcqa}
	\end{figure}
	
	In order to corroborate the discriminative ability of the metric $\widehat{d}_{CAQ}$ in this example, Figure \ref{curvesdcqa} displays the values of the sample mean as a function of the radius $r$ (blue curve). The dashed lines indicate the corresponding 5th and 95th sample quantiles. It seems that, in most cases, the differences between both time series are properly detected, although the degree of these differences dramatically varies with the choice of $r$, which highlights the importance of a proper selection of this parameter before carrying out distance computations. 
	
	\section{A fuzzy clustering algorithm for circular time series}\label{sectionfuzzyalgorithm}
	
	This section is devoted to introducing a fuzzy clustering algorithm for circular series based on the proposed distance $\widehat{d}_{CAQ}$.
	
	\subsection{A fuzzy $C$-medoids model based on the proposed dissimilarity}\label{subsectionfuzzyalgorithm}
	
	Consider a set of $n$ circular time series, $\mathbb{S}=\{\boldsymbol \Theta_{T_1}^{(1)}, \ldots, \boldsymbol \Theta_{T_n}^{(n)}\}$, where the $i$th series has length $T_i$. We want to perform fuzzy clustering on the elements of $\mathbb{S}$ in such a way that the series generated from similar stochastic processes are grouped together. To this aim, we propose to consider the fuzzy $C$-medoids clustering model introduced by \cite{krishnapuram1999fuzzy} in combination with the metric $\widehat{d}_{CAQ}$ proposed in Section~\ref{subsectiondissimilarity}. Thus, the goal is to determine the subset of $\mathbb{S}$ of size $C$, $\widetilde{\mathbb{S}}=\{\widetilde{\boldsymbol \Theta}^{(1)}, \ldots, \widetilde{\boldsymbol \Theta}^{(C)}\}$, whose elements are usually referred to as medoids, and the $n \times C$ matrix of fuzzy coefficients, $\boldsymbol U=(u_{ic})$, with $i=1, \ldots, n$ and $c=1, \ldots, C$, leading to the solution of the minimization problem 
	
	\begin{equation}\label{fcm}
		\min_{\widetilde{\mathbb{S}}, \boldsymbol U}\sum_{i=1}^{n}\sum_{c=1}^{C}u_{ic}^m\widehat{d}_{CQA}(i, c),  \, \, \,  \text{ with respect to} \, \, \sum_{c=1}^{C}u_{ic}=1, \, u_{ic} \ge 0,
	\end{equation}
	
	\noindent where $\widehat{d}_{CAQ}(i,c)=\widehat{d}_{CAQ}\big(\boldsymbol\theta_{T_i}^{(i)}, \widetilde{\boldsymbol\Theta}^{(c)}\big)$, $u_{ic} \in [0,1]$ represents the membership degree of the $i$th CTS in the $c$th cluster, and $m > 1$ is a real number, usually referred to as the fuzziness parameter, regulating the fuzziness of the partition. For $m=1$, the crisp version of the algorithm is obtained, so the solution takes the form $u_{ic}=1$ if the $i$th series pertains to cluster $c$ and $u_{ic}=0$ otherwise. As the value of $m$ increases, the boundaries between clusters get softer and the resulting partition is fuzzier. 
	
	The constrained optimization problem in \eqref{fcm} can be solved by means of the Lagrangian multipliers method, which leads to an iterative algorithm that alternately optimizes the membership degrees and the medoids. Specifically (see \cite{hoppner1999fuzzy}), the iterative solutions for the membership degrees are given by  
	\begin{equation}\label{updatemem}
		u_{ic}=\Bigg[\sum_{c'=1}^{C}\Bigg(\frac{\widehat{d}_{CAQ}(i, c)}{\widehat{d}_{CAQ}(i, c')}\Bigg)^{\frac{1}{m-1}}\Bigg]^{-1},
	\end{equation}
	for $i=1,\ldots,n$, and $c=1,\ldots, C$. 
	
	Once the membership degrees are obtained through \eqref{updatemem}, the $C$ series minimizing the objective function in \eqref{fcm} are selected as the new medoids. Specifically, for each $c \in \{1,\ldots,C\}$, the index $j_c$ satisfying
	
	\begin{equation}\label{jc}
		j_c = \argmin_{1 \leq j \leq n} \sum_{i=1}^{n} u_{ic}^m \widehat{d}_{CQA}\big(\boldsymbol\Theta_{T_i}^{(i)}, \boldsymbol\Theta_{T_j}^{(j)}\big),
	\end{equation}
	
	\noindent is computed. This two-step procedure is repeated until there is no change in the medoids or a maximum number of iterations is reached. An outline of the corresponding clustering algorithm is given in Algorithm~\ref{algorithm1}.

	\begin{algorithm}
		\textcolor{black}{\caption{\textcolor{black}{Fuzzy $C$-medoids algorithm based on the distance $\widehat{d}_{CQA}$. \label{algorithm1}}}}
		\begin{algorithmic}
			\State \textcolor{black}{Fix $C$, $m$ and \textit{max.iter}} 
			\State \textcolor{black}{Set $iter \, =0$}
			\State \textcolor{black}{Pick the initial medoids, $\widetilde{\mathbb{S}}=\{\widetilde{\boldsymbol \Theta}^{(1)}, \ldots, \widetilde{\boldsymbol \Theta}^{(C)}\}$}
			\Repeat
			\State \textcolor{black}{Set $\widetilde{\mathbb{S}}_{\text{OLD}}=\widetilde{\mathbb{S}}$}   
			\Comment{\textcolor{black}{Store the current medoids}}
			\State \textcolor{black}{Compute $u_{ic}$, $i=1,\ldots,n$, $c=1,\ldots,C$, using (\ref{updatemem})}
			\State \textcolor{black}{For each $c \in \{1,\ldots,C\}$, determine the index $j_c \in \{1,\ldots,n\}$ using \eqref{jc}}
			\State \textbf{return} \textcolor{black}{$\widetilde{X}_t^{(c)}=X_t^{(j_c)}$, for $c=1,\ldots,C$}  
			\Comment{\textcolor{black}{Update the medoids}}
			\State \textcolor{black}{$iter \, \gets iter \, + 1$}
			\Until{ \mbox{ \textcolor{black}{$\widetilde{\mathbb{S}}_{\text{OLD}}=\widetilde{\mathbb{S}} \mbox{ or } iter \, = \, max.iter$}} } 
			\State \textbf{return} \textcolor{black}{The final fuzzy partition and the corresponding set of medoids}
		\end{algorithmic}
	\end{algorithm}
	
	Two remarks related to the fuzzy $C$-medoids method described in Algorithm \ref{algorithm1} are given below. \\
	
	\begin{rem}
		\label{remadvantages}
		\textit{Advantages of the fuzzy $C$-medoids model}. The fuzzy $C$-medoids procedure outlined in Algorithm~\ref{algorithm1} allows us to identify a set of representative CTS belonging to the original collection, the medoids, whose overall distance to all other series in the set is minimal when the membership degrees with respect to a specific cluster are considered as weights (see the computation of $j_c$ in Algorithm~\ref{algorithm1}). As observed by \cite{kaufman2009finding}, it is often desirable that the prototypes synthesising the structural information of each cluster belong to the original data set, instead of obtaining ``virtual'' prototypes, as in the case of fuzzy $C$-means-based approaches \cite{dunn1973fuzzy, bezdek2013pattern}. For instance, the original set of series could be replaced by the set of medoids for exploratory purposes, thus substantially reducing the computational complexity of subsequent data mining tasks. The fuzzy $C$-medoids algorithm also exhibits classical advantages related to the fuzzy paradigm, including ability to produce richer clustering solutions than hard methods, identifying the vague nature of the prototypes, and the possibility of dealing with time series sharing different dynamic patterns, among others. 
	\end{rem}
	
	\vspace*{0.1cm}
	
	\begin{rem}
		\label{remhyperparameters}
		\textit{Hyperparameter selection in the fuzzy $C$-medoids model}. The fuzzy $C$-medoids procedure described in Algorithm~\ref{algorithm1} involves five hyperparameters, the number of clusters, $C$, the fuzziness parameter, $m$, the collection of lags, $\mathcal{L}$, the set of probability levels, $\mathcal{T}$, and the radius $r$. Concerning the set $\mathcal{L}$, we propose to select this set by using a mechanism analogous to the one described in Section 4.4 of \cite{lopez2023two}. Specifically, the procedure relies on assessing serial dependence at several lags for each CTS. To that aim, we employ the estimate $\widehat{\rho}_{JS}(l)$ along with its asymptotic distribution under the null hypothesis of zero circular autocorrelation (see Section 3 in \cite{jammalamadaka1988correlation}). Regarding the set $\mathcal{T}$, we propose to consider $\mathcal{T}=\{0.1, 0.5, 0.9\}$, since this set is usually enough to achieve a high accuracy when performing quantile-based time series clustering \cite{lafuente2016clustering,vilar2018quantile, lopez2021quantile, lopez2022quantile1}. Once the collections $\mathcal{L}$ and $\mathcal{T}$ have been selected, we propose to choose the remaining parameters ($C$, $m$ and $r$) by means of the following mechanism: (i) selecting a 3-dimensional grid of values for the corresponding hyperparameters, (ii) running the clustering algorithm for each element in the grid and recording the associated fuzzy partition, and (iii) selecting the element which lead to the minimum value of the Xie-Beni index \cite{xie1991validity} (the lower this index, the better the fuzzy partition in terms of compactness of the clusters and separation between clusters). It is worth highlighting that this kind of criteria based on internal indexes have been used in several works on time series clustering to perform hyperparameter selection \cite{lopez2021quantile, lopez2023two, lopez2023hard}.  
	\end{rem}
	
	\section{Simulation experiments}\label{sectionsimulations}
	
	In this section, we carry out a set of simulations with the aim of evaluating the behavior of the proposed algorithm in different scenarios of CTS clustering. First, we briefly describe some alternative dissimilarities that we consider for comparison purposes. Next, we explain how the performance of the algorithms is measured along with the corresponding simulation mechanism and results. 
	
	\subsection{Alternative metrics}
	
	To shed light on the behavior of the proposed clustering algorithm based on the distance $\widehat{d}_{CQA}$ (see Section \ref{sectionfuzzyalgorithm}), we decided to compare this method with fuzzy $C$-medoids procedures relying on alternative dissimilarities. Specifically, we considered the correlation-based metrics $\widehat{d}_{FL}$ and $\widehat{d}_{JS}$ introduced in Section \ref{subsectionmotivatingexample} and a metric based on the so-called quantile autocorrelation (QA). Given a real-valued, strictly stationary stochastic process $\{Y_t, t \in \mathbb{Z}\}$, with quantile function $q_Y$, and a couple of probability levels $\left(\tau, \tau^{\prime}\right) \in[0,1]^2$, QA of lag $l$ is defined as 
	
	\begin{equation}\label{dqa}
		\varphi(\tau, \tau', l)=\operatorname{Cov}\Big[I\left(Y_t\leq q_Y(\tau)\right), I\left(Y_{t+l} \leq q_Y(\tau')\right)\Big].
	\end{equation}
	
	By proceeding in the same way as with $\gamma(\tau, \tau', l, r)$ in Section \ref{subsectionmeasuring}, a natural estimate of $\varphi(\tau, \tau', l)$ can be defined. We denote this estimate by $\widehat{\varphi}(\tau, \tau', l)$. In a similar way, given two realizations $\boldsymbol Y_{T_1}^{(1)}=(y_1^{(1)}, y_2^{(1)}, \ldots, y_{T_1}^{(1)})$ and $\boldsymbol  Y_{T_2}^{(2)}=(y_1^{(2)}, y_2^{(2)}, \ldots, y_{T_2}^{(2)})$, from the processes $\{Y_t^{(1)}, t \in \mathbb{Z}\}$ and $\{Y_t^{(2)}, t \in \mathbb{Z}\}$, respectively, and the sets $\mathcal{L}=\{l_1, l_2, \ldots, l_L\}$ and $\mathcal{T}=\{\tau_1, \tau_2, \ldots, \tau_P\}$ of lags and probability levels, respectively, the corresponding QA-based distance is given by 
	
	\begin{equation}\label{estimateddcqa}
		\widehat{d}_{QA}\big(Y_t^{(1)}, Y_t^{(2)}\big)= \frac{1}{4LP^2}\sum_{k=1}^{L}\sum_{i=1}^{P}\sum_{j=1}^{P}\Big(\widehat{\varphi}(\tau_i, \tau_j, l_k)^{(1)}-\widehat{\varphi}(\tau_i, \tau_j, l_k)^{(2)}\Big)^2,
	\end{equation}
	
	\noindent where the superscripts (1) and (2) indicate that the corresponding estimates are computed from the realizations $\boldsymbol Y_{T_1}^{(1)}$ and $\boldsymbol  Y_{T_2}^{(2)}$, respectively. 
	
	The dissimilarity $\widehat{d}_{QA}$ is completely analogous to the distance based on the quantile autocovariance function proposed by \cite{lafuente2016clustering}, which exhibits a high ability to group time series generated from a broad range of dependence models both in the crisp \cite{lafuente2016clustering} and the fuzzy \cite{vilar2018quantile, lafuente2020robust} frameworks. Note that, unlike $\widehat{d}_{CQA}$, the metric $\widehat{d}_{QA}$ is not specifically designed to deal with time series taking values on the unit circle. Therefore, the performance of the latter distance is an essential benchmark for the former dissimilarity.

	\subsection{Experimental design and results}\label{subsectionedr}
	
	A comprehensive simulation study was carried out to evaluate the behavior of the fuzzy $C$-medoids algorithm based on the dissimilarity $\widehat{d}_{CQA}$. We attempted to consider an evaluation process allowing to draw general conclusions about the performance of the distance. To this aim, two different assessment schemes were designed. The first one includes scenarios with three different groups of CTS, and is aimed at evaluating the ability of the procedures to assign high (low) memberships if a given CTS pertains (not pertains) to a given group. The second one consists of scenarios constituted by two different groups of CTS plus one additional CTS which does not belong to any of the groups. We examine again the membership degrees of the series in the two clusters, but also that the isolated series is not placed in any of the clusters with a high membership degree. In this context, a given threshold is used to determine whether or not a membership degree in a given cluster is enough to assign the CTS to that group. 
	
	\subsubsection{First assessment scheme}\label{subsubsection1as}
	
	We considered three simple scenarios consisting of four three represented by the same type of generating processes, denoted by $\mathcal{C}_1$, $\mathcal{C}_2$ and $\mathcal{C}_3$. Each one of the groups contains five CTS, thus giving rise to a set of 15 CTS defining the true clustering partition. We attempted to construct scenarios with a wide variety of ordinal models commonly used in practice to deal with CTS. The generation mechanism is based on simulating realizations of a specific real-valued stochastic process, $\{X_t, t \in \mathbb{Z}\}$, and then constructing the corresponding circular time series by applying two transformations mapping the real line to the unit circle given by the functions $\eta_1$ and $\eta_2$ defined as
	
	\begin{equation}
		\eta_1(x)=x \mod(2\pi) \, \, \, \,\, \, \, \,\, \text{and} \, \, \, \,\, \, \, \,\, \eta_2(x)=2\arctan(x)+\pi,
	\end{equation} 
	
	\noindent where $\Phi$ denotes the distribution function of a standard normal random variable. Both functions $\eta_1$ and $\eta_2$ are a common choice for creating circular time series (see, e.g., Section 2 in \cite{fisher1994time}). 
	
	The generating models concerning the process $\{X_t, t \in \mathbb{Z}\}$ in each group are given below for each one of the scenarios. \\
	
	\noindent \textbf{Scenario 1}. Fuzzy clustering of CTS constructed from ARMA models \citealp{box2015time}. Let $\{X_t\}_{t \in \mathbb{Z}}$ be a stochastic process following the ARMA($p$, $q$)-type recursion given by
	
	\begin{equation}\label{arp}
		X_t=\sum_{i=1}^{p}\alpha_iX_{t-i}+\sum_{i=1}^{q}\beta_i\epsilon_{t-i}+\epsilon_t,
	\end{equation} 
	
	\noindent where $\alpha_1,\ldots, \alpha_p, \beta_1,\ldots, \beta_q$ are real numbers verifying the corresponding stationarity condition and $\{\epsilon_t\}_{t \in \mathbb{Z}}$ is a process formed by independent variables following the standard normal distribution. We fix $p=3$. The vector of coefficients $(\alpha_1, \alpha_2, \alpha_3, \beta_1, \beta_2, \beta_3)$ is set as indicated below.
	
	$$\begin{array}{l}
		\mathcal{C}_1:  \, \, \, (0.2, -0.2, 0.2, 0, 0, 0), \\
		\mathcal{C}_2:  \, \, \, (-0.2, 0.2, -0.2, 0, 0, 0), \\
		\mathcal{C}_3:  \, \, \, (0, 0, 0, 0.2, -0.2, 0.2).
	\end{array}$$ \\
	
	\noindent \textbf{Scenario 2}. Fuzzy clustering of CTS constructed from QAR models \cite{koenker2006quantile}. Let $\{X_t\}_{t \in \mathbb{Z}}$ be a stochastic process following the QAR($p$)-type recursion given by
	
	\begin{equation}\label{qar}
		X_t=f_0(U_t)+\sum_{i=1}^{p}f_i(U_t)X_{t-i},
	\end{equation} 
	
	\noindent where the $f_i$ are monotone real-valued functions with domain $[0,1]$ and $\{U_t\}_{t \in \mathbb{Z}}$ is a sequence of i.i.d. standard uniform random variables. We set $p=2$ and $f_0$ as the quantile function of a standard normal random variable. The functions $f_1$ and $f_2$ are fixed as indicated below.
	
	$$\begin{array}{l}
		\mathcal{C}_1:  \, \, \,   f_1=0.2(U_t-0.4), \, \, \, f_2=1.2(U_t-0.4), \\
		\mathcal{C}_2:  \, \, \,  f_1=-0.2(U_t-0.6), \, \, \, f_2=-1.2(U_t-0.6), \\
		\mathcal{C}_3:  \, \, \,  f_1=0, \, \, \, f_2=0.
	\end{array}$$
	
	Note that cluster $\mathcal{C}_3$ in the previous scenario is described by means of a white noise (WN) process. On the contrary, there exist nonnegative standard autocorrelations in the processes associated with clusters $\mathcal{C}_2$ and $\mathcal{C}_3$. \\
	
	\noindent \textbf{Scenario 3}. Fuzzy clustering of CTS constructed from GARCH models \cite{bollerslev1986generalized}. Let $\{X_t\}_{t \in \mathbb{Z}}$ be a stochastic process following the GARCH($p$, $q$)-type recursion given by
	
	\begin{equation}\label{garch}
		X_t=\sigma_t \epsilon_t, \quad \sigma_t^2=\alpha_0+\sum_{i=1}^p \alpha_i X_{t-i}^2+\sum_{j=1}^q \beta_j\sigma_{t-j}^2,
	\end{equation} 
	
	\noindent where $\alpha_0, \alpha_1,\ldots, \alpha_p, \beta_1,\ldots, \beta_q$ are real numbers verifying the corresponding stationarity condition and $\{\epsilon_t\}_{t \in \mathbb{Z}}$ is a process formed by independent variables following the standard normal distribution. We fix $p=q=2$. The vector of coefficients $(\alpha_0, \alpha_1, \alpha_2, \beta_1, \beta_2)$ is set as indicated below.
	
	$$\begin{array}{l}
		\mathcal{C}_1:  \, \, \, (0.1, 0.4, 0.4, 0.05, 0.05), \\
		\mathcal{C}_2:  \, \, \, (0.1, 0.05, 0.05, 0.4, 0.4), \\
		\mathcal{C}_3:  \, \, \, (0.1, 0.05, 0.4, 0.4, 0.05).
	\end{array}$$ \\

	As a preliminary step, a metric two-dimensional scaling (2DS) based on the metric  $\widehat{d}_{CQA}$ was carried out to gain insight about the capability of this distance to discriminate between the underlying groups. Given a distance matrix $\boldsymbol D=(D_{ij})_{1 \leq i,j \leq s}$, a 2DS finds the set of points $\{(a_i, b_i), i = 1,\ldots, n\}$ minimizing the loss function called stress given by  
	\begin{equation}\label{stress}
		\sqrt{\frac{\sum_{i \ne j=1}^{s}(\norm{(a_i, b_i)-(a_j, b_j)}-D_{ij})^2}{\sum_{i \ne j = 1}^{s}D_{ij}^2}}
	\end{equation}
	Thus, the goal is to represent the distances $D_{ij}$ in terms of Euclidean distances into a 2-dimensional space so that the original distances are preserved as well as possible. The lower the value of the stress function, the more reliable the 2DS configuration. This way, a 2DS plot provides a valuable visual representation of how the elements are located with respect to each other according to the original distance.
	
\begin{figure}
		\centering
		\includegraphics[width=1\textwidth]{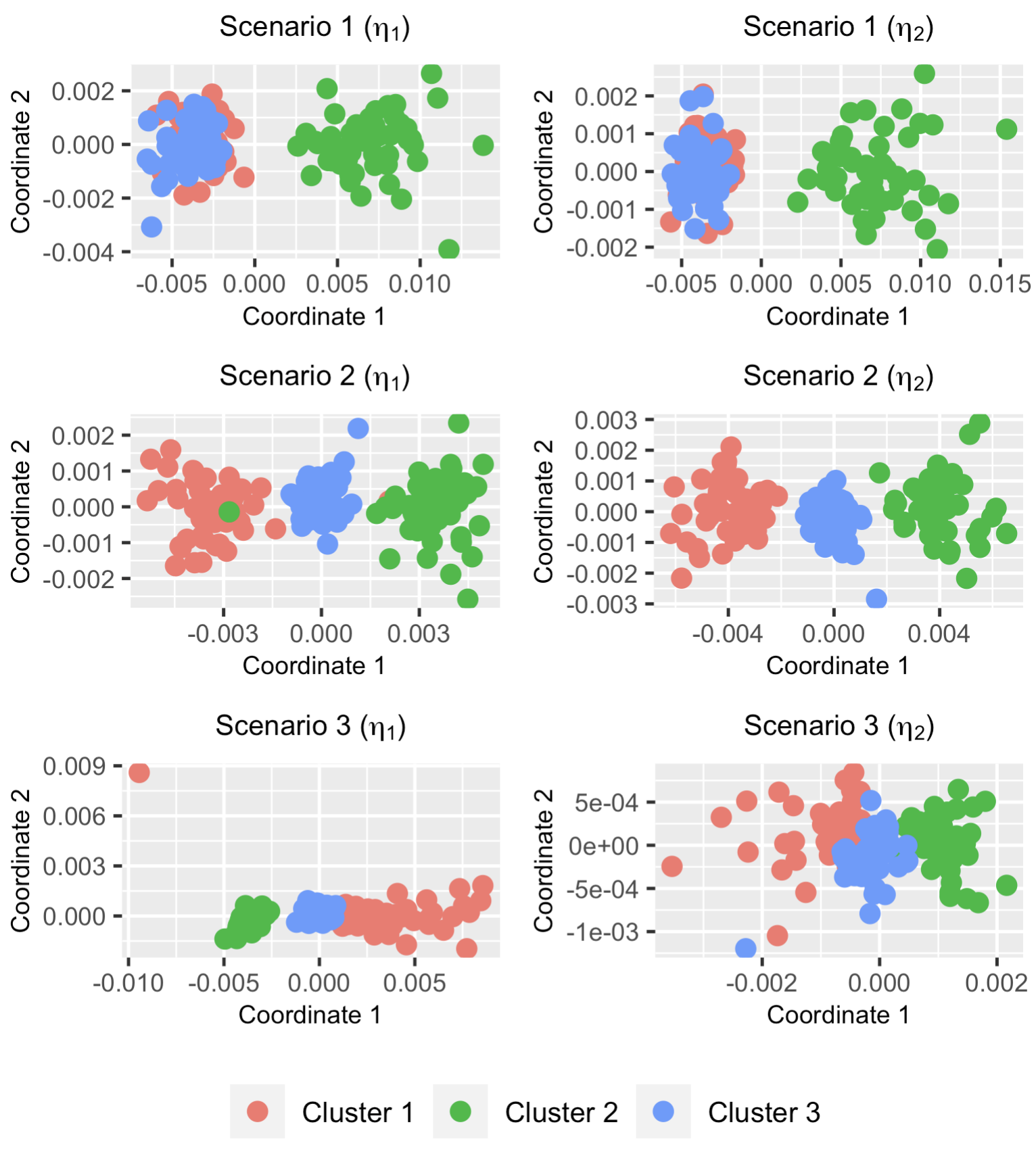}
		\caption{Two-dimensional scaling planes based on distance $\widehat{d}_{CQA}$ for Scenarios 1 ($T=500$), 2 ($T=500$) and 3 ($T=1000$) and two circular transformations.}
		\label{2ds}
	\end{figure}

	To obtain informative 2DS plots, 50 CTS of length $T = 500$ (Scenarios 1 and 2) and $T=1000$ (Scenario 3) from each generating model were simulated for each circular transformation. The 2DS was carried out for each set of 150 CTS by computing the pairwise dissimilarity matrices based on $\widehat{d}_{CQA}$. The parameter $r$ used for the computation of the metric was selected by considering the value which was chosen with the highest frequency in the clustering experiments shown throughout this section (see Remark \ref{remhyperparameters}). We considered the set of lags $\mathcal{L}=\{1, 2, 3\}$ in Scenario 1 and $\mathcal{L}=\{1, 2\}$ in Scenarios 2 and 3. The resulting plots are shown in Figure~\ref{2ds}, where a different colour was used for each generating process. It is worth highlighting that the $R^2$ value associated with the scaling is above 0.85 in all cases, thus concluding that the graphs in Figure~\ref{2ds} provide an accurate picture of the underlying representations according to both metrics.
	
	The reduced bivariate spaces in Figure \ref{2ds} show different configurations. In Scenario 1, the metric can clearly differentiate cluster $\mathcal{C}_2$ from clusters $\mathcal{C}_1$ and $\mathcal{C}_3$, but struggles to separate the latter groups. This is reasonable, since the dependence structures of the corresponding ARMA processes are very similar, and this property gets translated into the unit circle. On the contrary, $\widehat{d}_{CQA}$ can properly identify the three underlying groups in Scenarios 2 and 3, although some degree of overlap between clusters exists in the latter scenario due to the complex dependence patterns produced by the corresponding GARCH processes. In all cases, there are no substantial differences between the 2DS configurations associated with transformations $\eta_1$ and $\eta_2$. 
	
	The simulation study was carried out as follows. For each scenario, 5 CTS of length $T \in \{200, 500\}$ in Scenarios 1 and 2, and length $T \in \{500, 1000\}$ in Scenario 3 were generated from each process in order to execute the clustering algorithms twice and examine the effect of the series length. Several values of the fuzziness parameter $m$ were considered, namely $m \in \{1.2, 1.4, 1.6, 1.8, 2\}$. The problem of selecting a proper value for $m$ has been extensively addressed in the literature, although there seems to be no consensus about the optimal way of choosing this parameter (see the discussion in Section 3.1.6 of \cite{maharaj2011fuzzy}). When $m=1$, the hard version of the fuzzy $C$-medoids algorithm is obtained, while excessively large values of $m$ result in a partition with all memberships close to $1/C$, thus having a large degree of overlap between groups. As a consequence, selecting these values for $m$ is not recommended \cite{arabie1981overlapping}. Moreover, in the context of time series clustering, several works consider a grid of values for $m$ similar to our choice \cite{d2009autocorrelation,vilar2018quantile,lopez2021quantile, lopez2023two}. 
	
	Given a scenario and fixed values for $m$ and $T$, 200 simulations were executed. In each trial, the fuzzy $C$-medoids algorithm based on $\widehat{d}_{FL}$, $\widehat{d}_{JS}$, $\widehat{d}_{CQA}$ and $\widehat{d}_{QA}$ was applied with each value of $m$ as input. The number of clusters was set to $C=3$. The four dissimilarities were computed by employing the collection of lags $\mathcal{L}=\{1, 2, 3\}$ in Scenario~1 and $\mathcal{L}=\{1, 2\}$ in Scenarios 2 and 3, thus considering the maximum number of lags at each scenario. Concerning the distances $\widehat{d}_{CQA}$ and $\widehat{d}_{QA}$, the set of probability levels $\mathcal{T}=\{0.1, 0.5, 0.9\}$ was considered for their calculation. Moreover, the numerical experiments described throughout this section were also carried out by considering richer sets of probability levels, but no significant improvements were found. The parameter $r$ involved in the distance $\widehat{d}_{CQA}$ was selected by considering the procedure described in Remark \ref{remhyperparameters} and the fuzzy partition associated with the optimal value for $r$ was recorded. 
	
	In order to avoid the well-known issue of reaching local minima, the simulations were performed by considering the following two-step procedure:
	
	\begin{enumerate}
		\item  Running the corresponding fuzzy $C$-medoids algorithm by considering 200 different random starts for the set of initial medoids and storing the resulting membership matrix and medoids for each random start. 
		\item Among the 200 different clustering solutions, choosing the one giving rise to the minimum value of the objective function (see equation \eqref{fcm}).  
	\end{enumerate}
	
	Note that the idea of using several random initializations for the medoids and choosing the one leading to the optimal value of a given quantity (e.g., the objective function or a specific internal clustering quality index) has already been employed in some works on fuzzy clustering of time series (see, e.g., \cite{lopez2022quantile1, lopez2022spatial, lopez2023two}). In fact, this approach avoids the problem of having to choose a specific procedure for properly initializing the medoids. 
	
	Clustering acuraccy was assessed using the fuzzy extensions of the Adjusted Rand Index (ARIF) and the Jaccard Index (JIF) introduced by \cite{campello2007fuzzy}. Both indexes are obtained by reformulating the original ones in terms of the fuzzy set theory, which allows to compare the true (hard) partition with an experimental fuzzy partition. ARIF and JIF take values in the intervals $[-1, 1]$ and $[0, 1]$, respectively, with values closer to 1 indicating a more accurate clustering solution. 
	
	
	The average values of ARIF and JIF based on the 200 simulation trials are shown in Table~\ref{tablet1} for the transformation $\eta_1$ and in Table~\ref{tablet2} for the transformation $\eta_2$. We can see that, in both cases, all distances decrease their performance when increasing the value of $m$. This is reasonable and expected, since larger values of $m$ produce a smoother boundary between the four well-separated clusters, thus making the classification fuzzier and resulting in lower values of ARIF and JIF. 
	
	\begin{table}
		\begin{tabular}{cccccc|cccccc}
			\hline \multirow{3}{*}{ Scenario 1} & & \multicolumn{4}{c}{ARIF} & \multicolumn{4}{c|}{JIF} \\
			\hline & & & & & & & & &  \\
			&  & $\widehat{d}_{FL}$ & $\widehat{d}_{JS}$ & $\widehat{d}_{CQA}$ & $\widehat{d}_{QA}$ & $\widehat{d}_{FL}$ & $\widehat{d}_{JS}$ & $\widehat{d}_{CQA}$ & $\widehat{d}_{QA}$ \\
			\hline \multirow[t]{5}{*}{$T=200$} & $m=1.2$  & \textbf{0.60} & \textbf{0.60} & 0.53 & 0.38 & \textbf{0.57} & \textbf{0.57} & 0.52 & 0.42 \\ 
			\hline & $m=1.4$ & \textbf{0.56} & \textbf{0.56} & 0.46 & 0.27 & \textbf{0.54} & \textbf{0.54} & 0.47 & 0.36 \\ 
			\hline & $m=1.6$  & \textbf{0.53} & \textbf{0.53} & 0.39 & 0.20 & \textbf{0.52} & \textbf{0.52} & 0.43 & 0.32 \\
			\hline & $m=1.8$  & \textbf{0.49} & \textbf{0.48} & 0.32 & 0.15 & \textbf{0.49} & \textbf{0.49} & 0.39 & 0.29 \\ 
			\hline & $m=2.0$  & \textbf{0.45} & \textbf{0.45} & 0.30 & 0.13 & \textbf{0.47} & \textbf{0.46} & 0.38 & 0.28 \\ 
			\hline \multirow[t]{5}{*}{$T=500$} & $m=1.2$ & \textbf{0.63} & \textbf{0.63} & 0.58 & 0.52 & \textbf{0.60} & \textbf{0.60} & 0.56 & 0.52 \\ 
			\hline & $m=1.4$ & \textbf{0.60} & \textbf{0.59} & 0.52 & 0.42 & \textbf{0.57} & \textbf{0.57} & 0.52 & 0.45 \\ 
			\hline & $m=1.6$  & \textbf{0.56} & \textbf{0.56} & 0.46 & 0.33 & \textbf{0.55} & \textbf{0.55} & 0.48 & 0.40 \\ 
			\hline & $m=1.8$ & \textbf{0.53} & \textbf{0.53} & 0.41 & 0.27 & \textbf{0.53} & \textbf{0.53} & 0.45 & 0.36 \\ 
			\hline & $m=2.0$ & \textbf{0.51 }& \textbf{0.51} & 0.35 & 0.23 & \textbf{0.51} & \textbf{0.51} & 0.41 & 0.34 \\ 
			\hline \multicolumn{1}{c}{Scenario 2} \\
			& & $\widehat{d}_{FL}$ & $\widehat{d}_{JS}$ & $\widehat{d}_{CQA}$ & $\widehat{d}_{QA}$ & $\widehat{d}_{FL}$ & $\widehat{d}_{JS}$ & $\widehat{d}_{CQA}$ & $\widehat{d}_{QA}$ \\
			\hline \multirow[t]{5}{*}{$T=200$} & $m=1.2$  & 0.27 & 0.24 & \textbf{0.82} & 0.34 & 0.35 & 0.33 & \textbf{0.79} & 0.39 \\ 
			\hline & $m=1.4$ & 0.22 & 0.20 & \textbf{0.62} & 0.23 & 0.32 & 0.31 & \textbf{0.60} & 0.33 \\
			\hline & $m=1.6$  & 0.18 & 0.17 & \textbf{0.50} & 0.16 & 0.30 & 0.29 & \textbf{0.50} & 0.29 \\
			\hline & $m=1.8$  & 0.16 & 0.15 & \textbf{0.41} & 0.13 & 0.29 & 0.28 & \textbf{0.44} & 0.28 \\ 
			\hline & $m=2.0$  & 0.13 & 0.12 & \textbf{0.33} & 0.10 & 0.28 & 0.27 & \textbf{0.39} & 0.26 \\ 
			\hline \multirow[t]{5}{*}{$T=500$} & $m=1.2$  & 0.37 & 0.34 & \textbf{0.99} & 0.58 & 0.41 & 0.39 & \textbf{0.98} & 0.56 \\ 
			\hline & $m=1.4$ & 0.32 & 0.29 & \textbf{0.91} & 0.40 & 0.38 & 0.36 & \textbf{0.88} & 0.43 \\ 
			\hline & $m=1.6$  & 0.28 & 0.25 & \textbf{0.78} & 0.31 & 0.36 & 0.34 & \textbf{0.74} & 0.38 \\ 
			\hline & $m=1.8$ & 0.24 & 0.21 & \textbf{0.66} & 0.24 & 0.33 & 0.32 & \textbf{0.62} & 0.34 \\
			\hline & $m=2.0$ & 0.20 & 0.17 & \textbf{0.57} & 0.20 & 0.31 & 0.30 & \textbf{0.55} & 0.31 \\ 
			\hline \multicolumn{1}{c}{Scenario 3} \\
			& & $\widehat{d}_{FL}$ & $\widehat{d}_{JS}$ & $\widehat{d}_{CQA}$ & $\widehat{d}_{QA}$ & $\widehat{d}_{FL}$ & $\widehat{d}_{JS}$ & $\widehat{d}_{CQA}$ & $\widehat{d}_{QA}$ \\
			\hline \multirow[t]{5}{*}{$T=500$} & $m=1.2$ & 0.17 & 0.16 & \textbf{0.77} & 0.17 & 0.30 & 0.29 & \textbf{0.73}& 0.29 \\
			\hline & $m=1.4$  & 0.14 & 0.12 & \textbf{0.60} & 0.11 & 0.28 & 0.27 & \textbf{0.58} & 0.26 \\ 
			\hline & $m=1.6$ & 0.12 & 0.11 & \textbf{0.48} & 0.08 & 0.27 & 0.26 & \textbf{0.49} & 0.25 \\ 
			\hline & $m=1.8$ & 0.10 & 0.09 & \textbf{0.41} & 0.06 & 0.26 & 0.25 & \textbf{0.44} & 0.25 \\ 
			\hline & $m=2.0$  & 0.09 & 0.08 & \textbf{0.35} & 0.06 & 0.26 & 0.25 & \textbf{0.40} & 0.24 \\
			\hline \multirow[t]{5}{*}{$T=1000$} & $m=1.2$ & 0.16 & 0.15 & \textbf{0.96} & 0.21 & 0.29 & 0.29 & \textbf{0.95}& 0.32 \\ 
			\hline & $m=1.4$  & 0.13 & 0.12 & \textbf{0.85} & 0.13 & 0.28 & 0.27 & \textbf{0.82} & 0.28 \\ 
			\hline & $m=1.6$ & 0.11 & 0.09 & \textbf{0.74}& 0.10 & 0.27 & 0.26 & \textbf{0.70} & 0.26 \\ 
			\hline & $m=1.8$  & 0.10 & 0.09 & \textbf{0.63} & 0.08 & 0.26 & 0.26 & \textbf{0.60} & 0.25 \\ 
			\hline & $m=2.0$  & 0.09 & 0.08 & \textbf{0.54} & 0.07 & 0.26 & 0.25 & \textbf{0.53} & 0.25 \\
			\hline
		\end{tabular}
		\caption{Average values of ARIF and JIF obtained by the fuzzy $C$-medoids clustering algorithm based on several dissimilarities. Scenarios 1, 2 and 3. For each value of $m$ and $T$, the best result is shown in bold. The corresponding CTS were constructed using the transformation $\eta_1(x)=x \mod(2\pi)$.}
		\label{tablet1}
	\end{table}

	\begin{table}
		\begin{tabular}{cccccc|cccccc}
			\hline \multirow{3}{*}{ Scenario 1} & & \multicolumn{4}{c}{ARIF} & \multicolumn{4}{c|}{JIF} \\
			\hline & & & & & & & & &  \\
			&  & $\widehat{d}_{FL}$ & $\widehat{d}_{JS}$ & $\widehat{d}_{CQA}$ & $\widehat{d}_{QA}$ & $\widehat{d}_{FL}$ & $\widehat{d}_{JS}$ & $\widehat{d}_{CQA}$ & $\widehat{d}_{QA}$ \\
			\hline \multirow[t]{5}{*}{$T=200$} & $m=1.2$  & 0.58 & \textbf{0.59} & 0.54 & 0.56 & 0.56 & \textbf{0.57} & 0.53 & 0.55 \\ 
			\hline & $m=1.4$  & 0.54 & \textbf{0.56} & 0.45 & 0.49 & 0.53 & \textbf{0.55} & 0.47 & 0.50 \\
			\hline & $m=1.6$   & 0.49 & \textbf{0.52} & 0.38 & 0.42 & 0.50 & \textbf{0.52} & 0.43 & 0.45 \\ 
			\hline & $m=1.8$  & 0.44 & \textbf{0.48} & 0.33 & 0.37 & 0.46 & \textbf{0.49} & 0.40 & 0.42 \\ 
			\hline & $m=2.0$  & 0.40 & \textbf{0.44} & 0.29 & 0.32 & 0.43 & \textbf{0.46} & 0.37 & 0.39 \\ 
			\hline \multirow[t]{5}{*}{$T=500$} & $m=1.2$ & \textbf{0.62} & \textbf{0.62} & 0.57 & 0.58 & 0.59 & \textbf{0.60} & 0.55 & 0.57  \\
			\hline & $m=1.4$ & \textbf{0.58} & \textbf{0.58} & 0.49 & 0.52 & \textbf{0.56} & \textbf{0.56} & 0.50 & 0.52 \\
			\hline & $m=1.6$  & 0.53 & \textbf{0.55} & 0.44 & 0.48 & 0.53 & \textbf{0.54} & 0.46 & 0.49 \\ 
			\hline & $m=1.8$ & 0.51 & \textbf{0.53} & 0.41 & 0.45 & 0.51 & \textbf{0.52} & 0.44 & 0.47 \\ 
			\hline & $m=2.0$  & 0.47 & \textbf{0.50} & 0.37 & 0.40 & 0.48 & \textbf{0.50} & 0.42 & 0.44 \\ 
			\hline \multicolumn{1}{c}{Scenario 2} \\
			& & $\widehat{d}_{FL}$ & $\widehat{d}_{JS}$ & $\widehat{d}_{CQA}$ & $\widehat{d}_{QA}$ & $\widehat{d}_{FL}$ & $\widehat{d}_{JS}$ & $\widehat{d}_{CQA}$ & $\widehat{d}_{QA}$ \\
			\hline \multirow[t]{5}{*}{$T=200$} & $m=1.2$  & 0.32 & 0.22 & \textbf{0.89} & 0.84 & 0.38 & 0.32 & \textbf{0.87} & 0.80 \\
			\hline & $m=1.4$ & 0.26 & 0.19 & \textbf{0.73} & 0.63 & 0.34 & 0.30 & \textbf{0.69} & 0.60 \\
			\hline & $m=1.6$  & 0.21 & 0.15 & \textbf{0.58} & 0.49 & 0.32 & 0.28 & \textbf{0.56} & 0.50 \\
			\hline & $m=1.8$ & 0.20 & 0.14 & \textbf{0.49} & 0.40 & 0.31 & 0.28 & \textbf{0.49} & 0.43 \\
			\hline & $m=2.0$  & 0.18 & 0.12 & \textbf{0.41 }& 0.33 & 0.30 & 0.27 & \textbf{0.44} & 0.39 \\ 
			\hline \multirow[t]{5}{*}{$T=500$} & $m=1.2$ & 0.44 & 0.32 & \textbf{1.00} & 0.99 & 0.46 & 0.37 & \textbf{1.00} & 0.98 \\ 
			\hline & $m=1.4$ & 0.39 & 0.27 & \textbf{0.95} & 0.90 & 0.42 & 0.35 & \textbf{0.93} & 0.87 \\ 
			\hline & $m=1.6$ & 0.34 & 0.23 & \textbf{0.84} & 0.77 & 0.39 & 0.33 & \textbf{0.80} & 0.72 \\ 
			\hline & $m=1.8$  & 0.30 & 0.19 & \textbf{0.73} & 0.65 & 0.37 & 0.30 & \textbf{0.69} & 0.61 \\ 
			\hline & $m=2.0$ & 0.27 & 0.17 & \textbf{0.64} & 0.55 & 0.35 & 0.30 & \textbf{0.60} & 0.54 \\ 
			\hline \multicolumn{1}{c}{Scenario 3} \\
			& & $\widehat{d}_{FL}$ & $\widehat{d}_{JS}$ & $\widehat{d}_{CQA}$ & $\widehat{d}_{QA}$ & $\widehat{d}_{FL}$ & $\widehat{d}_{JS}$ & $\widehat{d}_{CQA}$ & $\widehat{d}_{QA}$ \\
			\hline \multirow[t]{5}{*}{$T=500$} & $m=1.2$  & 0.22 & 0.16 & \textbf{0.59} & 0.49 & 0.32 & 0.29 & \textbf{0.58} & 0.49 \\
			\hline & $m=1.4$  & 0.18 & 0.12 & \textbf{0.43} & 0.35 & 0.30 & 0.27 &\textbf{0.45} & 0.40 \\ 
			\hline & $m=1.6$ & 0.15 & 0.09 & \textbf{0.33} & 0.27 & 0.29 & 0.26 & \textbf{0.39} & 0.35 \\ 
			\hline & $m=1.8$ & 0.13 & 0.09 & \textbf{0.28} & 0.21 & 0.28 & 0.25 & \textbf{0.36} & 0.32 \\
			\hline & $m=2.0$ & 0.12 & 0.07 & \textbf{0.24} & 0.18 & 0.27 & 0.25 & \textbf{0.34}& 0.31 \\ 
			\hline \multirow[t]{5}{*}{$T=1000$} & $m=1.2$ & 0.22 & 0.15 & \textbf{0.83} & 0.67 & 0.32 & 0.28 & \textbf{0.80} & 0.64 \\ 
			\hline & $m=1.4$  & 0.19 & 0.13 & \textbf{0.68} & 0.51 & 0.30 & 0.27 & \textbf{0.65} & 0.51 \\ 
			\hline & $m=1.6$ & 0.16 & 0.10 & \textbf{0.54} & 0.40 & 0.29 & 0.26 & \textbf{0.53} & 0.44 \\
			\hline & $m=1.8$  & 0.12 & 0.08 & \textbf{0.45} & 0.33 & 0.27 & 0.25 & \textbf{0.46} & 0.39 \\ 
			\hline & $m=2.0$  & 0.12 & 0.07 & \textbf{0.37}& 0.27 & 0.27 & 0.25 & \textbf{0.42} & 0.36 \\ 
			\hline
		\end{tabular}
		\caption{Average values of ARIF and JIF obtained by the fuzzy $C$-medoids clustering algorithm based on several dissimilarities. Scenarios 1, 2 and 3. For each value of $m$ and $T$, the best result is shown in bold. The corresponding CTS were constructed using the transformation $\eta_2(x)=2 \arctan(x)+1$.}
		\label{tablet2}
	\end{table}
	
	According to Table \ref{tablet1}, the metrics based on circular autocorrelations ($\widehat{d}_{FL}$ and $\widehat{d}_{JS}$) achieve the best clustering accuracy in Scenario 1, with no significant differences existing between them. However, these dissimilarities are far from reaching the true clustering partition even when $T=500$ and $m=1.2$. This is due to the similarity of processes associated with clusters $\mathcal{C}_1$ and $\mathcal{C}_3$ in terms of dependence structures. In fact, the value of the estimated features $\widehat{\rho}_{FL}(1)$, $\widehat{\rho}_{FL}(2)$ and $\widehat{\rho}_{FL}(3)$ (and $\widehat{\rho}_{JS}(1)$, $\widehat{\rho}_{JS}(2)$ and $\widehat{\rho}_{JS}(3)$) are very similar for the series generated from both processes, thus producing clustering partitions with a large degree of overlap between clusters $\mathcal{C}_1$ and $\mathcal{C}_3$. The proposed metric $\widehat{d}_{CQA}$ shows only a slightly worse clustering accuracy than $\widehat{d}_{FL}$ and $\widehat{d}_{JS}$, specially for low values of $m$. On the contrary, this metric dramatically outperforms the three alternative distances in terms of clustering effectiveness in Scenarios 2 and 3. In these scenarios, the transformed QAR and GARCH processes produce very complex forms of circular dependence that can not be detected neither by measures of circular autocorrelation nor by the noncircular feature QA. However, the different dependence patterns become clearly visible when employing the quantity CQA through the distance $\widehat{d}_{CQA}$.
	
	The results in Table \ref{tablet2} (transformation $\eta_2$) are very similar to the ones in Table \ref{tablet1} for metrics $\widehat{d}_{FL}$, $\widehat{d}_{JS}$ and $\widehat{d}_{CQA}$, with the former distances getting the best clustering accuracy in Scenario 1, and the latter one outperforming the alternative metrics in Scenarios 2 and 3. However, metric $\widehat{d}_{QA}$ substantially increases its results when transformation $\eta_2$ is used. This is because this transformation gives rise to circular processes whose serial dependence patterns are moderately different in terms of QA, even though this feature is aimed at analyzing time series without a circular character. In any case, $\widehat{d}_{QA}$ achieves slightly higher scores than $\widehat{d}_{CQA}$ in Scenarios 2 and 3, thus corroborating the discriminative ability of the CQA measure even when the underlying processes can be identified by means of classical quantile-based measures of serial dependence. 
	
	Overall, the above experiments showed the great behavior of $\widehat{d}_{CQA}$ to perform fuzzy clustering of CTS when the true partition is formed by well-separated clusters. The superiority of this metric with respect to distances based on linear measures of circular dependence ($\widehat{d}_{FL}$ and $\widehat{d}_{JS}$), and metrics ignoring the circular nature of the series ($\widehat{d}_{QA}$) was corroborated in scenarios characterized by several types of circular processes with different dependence patterns. This highlights the importance of constructing distances specifically designed to deal with circular time series.
	
	\subsubsection{Second assessment scheme}\label{subsubsection2as}
	
	A second simulation experiment was conducted to analyze the effect of ambiguous series, whose presence introduces certain degree of uncertainty and increases the fuzzy nature of the clustering task. Three new scenarios consisting of two well-separated clusters of 5 CTS each and a single isolated series arising from a different process are defined as follows. \\
	
	\vspace*{0.2cm}
	
	\noindent \textbf{Scenario 4}. A set of 11 OTS, where five series are generated as in cluster $\mathcal{C}_1$ of Scenario 1, five additional series are generated as in cluster $\mathcal{C}_2$ of Scenario 1, and one isolated series is generated from a white noise (WN) process. 
	
	\vspace*{0.2cm}
	
	\noindent \textbf{Scenario 5}. A set of 11 OTS, where five series are generated as in cluster $\mathcal{C}_1$ of Scenario 2, five additional series are generated as in cluster $\mathcal{C}_2$ of Scenario 2, and one isolated series is generated from a WN process. 
	
	\vspace*{0.2cm}
	
	\noindent \textbf{Scenario 6}. A set of 11 OTS, where five series are generated as in cluster $\mathcal{C}_1$ of Scenario 3, five additional series are generated as in cluster $\mathcal{C}_2$ of Scenario 3, and one isolated series is generated from a GARCH($2$, $2$) process with vector of coefficients $(\alpha_0, \alpha_1, \alpha_2, \beta_1, \beta_2)=(0.1, 0.225, 0.225, 0.225, 0.225)$.  \\
	
	Note that Scenarios~4, 5 and 6 have been designed in a way that the isolated series is expected to lay ``in the middle'' of both clusters. In other words, a metric capable of discriminating between circular generating processes should be able to produce similar distance values from the isolated series to series from cluster $\mathcal{C}_1$ and cluster $\mathcal{C}_2$ indistinctly. 
	
	The values for $T$, the number of simulation trials and the number of random starts for the set of medoids, as well as the sets $\mathcal{L}$ and $\mathcal{T}$, were fixed as in Scenarios~1--3. The number of clusters was set to $C=2$. Assessment was performed in a different way. We computed the proportion of times that: the five series from cluster $\mathcal{C}_1$ grouped together in one group, the five series from cluster $\mathcal{C}_2$ clustered together in another group, and the isolated series had a relatively high membership degree with respect to each of the groups. To this aim, a cutoff point must be determined to conclude when a series is assigned to a specific cluster. We decided to use the cutoff value of 0.7, i.e. the $i$th OTS was placed into the $c$th cluster if $u_{ic} > 0.7$. On the contrary, a time series was considered to simultaneously belong to both clusters if its membership degrees were both below 0.7. The use of a cutoff value to assess fuzzy clustering algorithms has already been considered in prior works \cite{maharaj2011fuzzy, d2012wavelets, lopez2022quantile1, lopez2023two}. Specifically, some arguments for this choice are given in \cite{maharaj2011fuzzy}.
	
\begin{figure}
		\centering
		\includegraphics[width=1\textwidth]{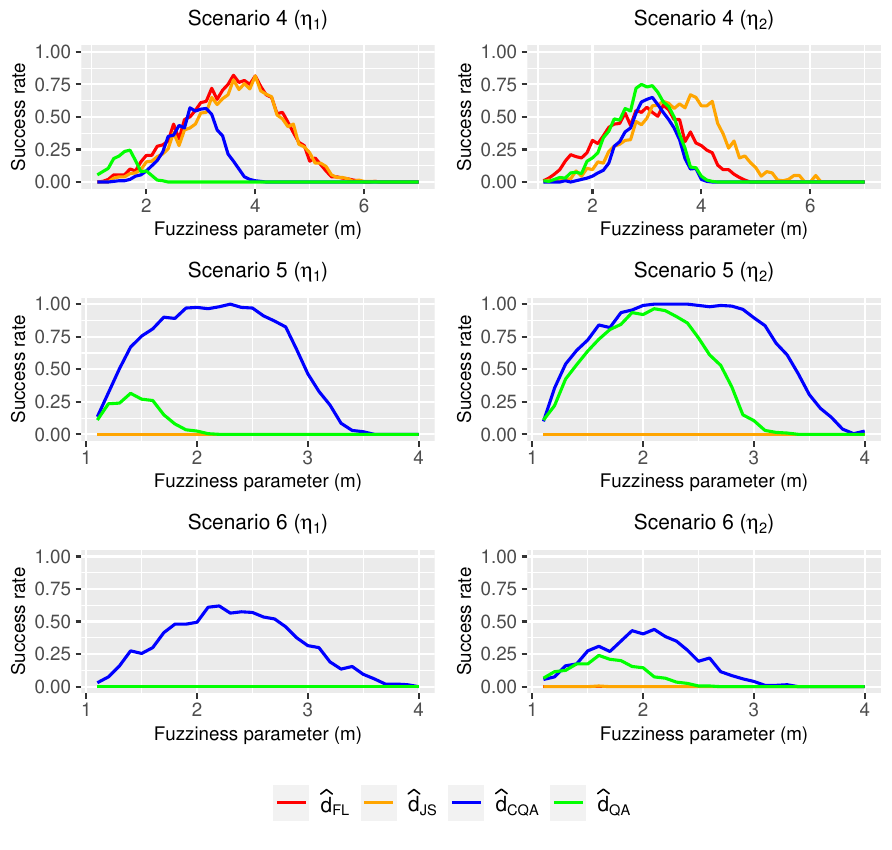}
		\caption{Rates of correct classification as function of $m$ obtained by the fuzzy $C$-medoids clustering algorithm based on several dissimilarities with a cutoff of 0.7. Scenarios 4 ($T=500$), 5 ($T=500$) and 6 ($T=1000$).}
		\label{curvesscenarios456}
	\end{figure}

	Note that this evaluation criterion is very sensitive to the selection of $m$, since a single series with membership degrees failing to fulfil the required condition results in an incorrect classification. In fact, the different metrics could achieve their best behaviour for rather different values of $m$. For this reason, we decided to run the clustering algorithms for a grid of values for $m$ on the interval $(1, 4]$. Figure~\ref{curvesscenarios456} contains the corresponding curves of rates of correct classification as a function of $m$ for the largest values of $T$. 
	
	The graphs in Figure \ref{curvesscenarios456} confirm that the fuzziness parameter dramatically affects the clustering performance. In all cases, low and high values of $m$ produce poor rates of correct classification, since partitions with all memberships close to 1 or to $1/2$ are respectively returned, thus resulting in failed trials. On the contrary, moderate values of $m$ generally result in higher clustering effectiveness, although the optimal range varies for each metric. In Scenario 4, the proposed distance gets worse correct classification rates than the metrics based on circular autocorrelations ($\widehat{d}_{FL}$ and $\widehat{d}_{JS}$), and even than its noncircular counterpart ($\widehat{d}_{QA}$) when transformation $\eta_2$ is used. In contrast, metric $\widehat{d}_{CQA}$ dramatically outperforms the alternative distances in Scenarios 5 and 6. In particular, distances $\widehat{d}_{FL}$ and $\widehat{d}_{JS}$ are not able to achieve a single correct classification in these scenarios, while $\widehat{d}_{QA}$ improves their performance when transformation $\eta_2$ is used, which is coherent with the results in Tables \ref{tablet1} and \ref{tablet2}. In all cases, increasing the series length results in better rates of correct classification. Note that the previous results demonstrate the importance of a suitable selection of $m$, although this issue is not addressed here because there are several procedures available in the literature for this purpose. 
	
	Rigorous comparisons based on Figure \ref{curvesscenarios456} can be made by computing: (i) the maximum value of each curve, and (ii) the area under each curve, denoted by AUFC (area under the fuzziness curve), which was already used by \cite{lopez2022quantile1, lopez2023two}. The values for both quantities are given in Table \ref{tablescenarios67}, and clearly corroborate the great performance of metrics $\widehat{d}_{FL}$ and $\widehat{d}_{JS}$ in Scenario 4, and of $\widehat{d}_{CQA}$ in Scenarios 5 and 6. In fact, in these scenarios, the proposed metric significantly outperforms $\widehat{d}_{QA}$ according to AUFC  even when transformation $\eta_2$ is used. 
	
	\begin{table}
		\centering
		\resizebox{11cm}{!}{\begin{tabular}{cccccc|cccc}  \hline
				Scenario 4	&  & $\eta_1$ & & & & $\eta_2$ & & &  \\  
				& & $\widehat{d}_{FL}$  &  $\widehat{d}_{JS}$ &  $\widehat{d}_{CQA}$ & $\widehat{d}_{QA}$ & $\widehat{d}_{FL}$  &  $\widehat{d}_{JS}$ &  $\widehat{d}_{CQA}$ &   $\widehat{d}_{QA}$  \\  \hline
				$T=200$	  & Maximum & \textbf{0.490} & 0.485 &  0.215 & 0.400 & 0.300 & 0.380 & 0.300 & \textbf{0.470}    \\ 
				& AUFC & \textbf{0.864} & 0.789 & 0.195 & 0.014 & 0.411 & \textbf{0.631} & 0.235 & 0.458 \\ \hline 
				$T=500$	  & Maximum & \textbf{0.820} & 0.815 &  0.570 & 0.245 & 0.590 & 0.650 & 0.670 & \textbf{0.750} \\ 
				& AUFC & \textbf{1.736} & 1.592 & 0.638 & 0.148 & 1.140 & \textbf{1.323} & 0.700 & 0.992  \\ \hline 
				Scenario 5	&  & $\eta_1$ & & & & $\eta_2$ & & &  \\  
				& & $\widehat{d}_{FL}$  &  $\widehat{d}_{JS}$ &  $\widehat{d}_{CQA}$ &   $\widehat{d}_{QA}$ & $\widehat{d}_{FL}$  &  $\widehat{d}_{JS}$ &  $\widehat{d}_{CQA}$ &   $\widehat{d}_{QA}$  \\  \hline
				$T=200$	  & Maximum &  0.005 & 0.000 &  \textbf{0.680} & 0.035 & 0.000 & 0.000 & \textbf{0.845} &  0.560  \\ 
				& AUFC & 0.001 & 0.000 & \textbf{0.571} & 0.007  & 0.000 & 0.000 & \textbf{0.899} &  0.378  \\ \hline 
				$T=500$	  & Maximum & 0.000 & 0.000 &  \textbf{1.000} & 0.315 & 0.000 & 0.000 & \textbf{1.000} & 0.965   \\ 
				& AUFC & 0.000 & 0.000 & \textbf{1.616} & 0.167 & 0.000 & 0.000 & \textbf{1.999} & 1.234   \\ \hline 
				Scenario 6	&  & $\eta_1$ & & & & $\eta_2$ & & &  \\  
				& & $\widehat{d}_{FL}$  &  $\widehat{d}_{JS}$ &  $\widehat{d}_{CQA}$ &   $\widehat{d}_{QA}$ & $\widehat{d}_{FL}$  &  $\widehat{d}_{JS}$ &  $\widehat{d}_{CQA}$ &   $\widehat{d}_{QA}$  \\  \hline
				$T=500$	  & Maximum & 0.005 & 0.000 & \textbf{0.260} & 0.000 & 0.000 & 0.000 & \textbf{0.190} & 0.095   \\
				& AUFC & 0.005 & 0.000 & \textbf{0.249} & 0.000 & 0.000 & 0.000 & \textbf{0.118} & 0.040  \\ \hline
				$T=1000$	  & Maximum & 0.000 & 0.000 & \textbf{0.620} & 0.000 & 0.000 & 0.005 & \textbf{0.440} & 0.240   \\ 
				& AUFC & 0.000 & 0.000 & \textbf{0.905} & 0.000 & 0.000 & 0.005 & \textbf{0.468} & 0.178 \\ \hline 
		\end{tabular}}
		\caption{Maximum rates of correct classification and AUFC obtained by the fuzzy $C$-medoids clustering algorithm based on several distances for a cutoff value of 0.7. Scenarios 4, 5 and 6. The best results are shown in bold.}
		\label{tablescenarios67}
	\end{table}
	
	In sum, the experiments shown in this section showed the clustering effectiveness of the proposed dissimilarity when dealing with time series databases subject to a certain degree of uncertainty. Furthermore, the higher values of the AUFC attained by $\widehat{d}_{CQA}$ with respect to $\widehat{d}_{QA}$ in Scenarios 5 and 6 indicate a greater robustness of the former distance to the choice of $m$. This is a nice property, since the optimal selection of the fuzziness parameter is still an open problem in the fuzzy clustering literature. 
	
	\section{Application. Fuzzy clustering of wind direction in Saudi Arabia}\label{sectionapplications}
	
	This section is devoted to showing two interesting applications of the proposed clustering procedure involving wind data in Saudi Arabia. In both cases, we first describe the considered data along with some exploratory analyses and, afterwards, we show the results of applying the clustering algorithms. 
	
	\subsection{Fuzzy clustering of wind direction in a particular location}\label{subsectionapplication1}
	
	We considered a dataset of hourly time series of wind direction measured in several locations of Saudi Arabia between the years 2010 and 2017. This database was sourced from the KAPSARC Energy Data Portal\footnote{\href{https://datasource.kapsarc.org/explore/dataset/saudi-hourly-weather-data/information/}{https://datasource.kapsarc.org/explore/dataset/saudi-hourly-weather-data/information/}}. Specifically, we retained the hourly observations related to the months of January, February, March and December (winter months), and June, July, August and September (summer months), recorded in the city of Abha (located in southwestern region of the country). We treat each time series as monthly data from the defined winter and summer months over the eight years, 2010-2017, resulting in a dataset containing 64 time series. Note that, from a meteorological point of view, it is reasonable to assume that the temporal behavior of the wind direction depends on the season of the year. Moreover, as we are considering a fuzzy approach, we expect to identify some series showing a vague behavior. The fuzzy $C$-medoids algorithm based on the distance $\widehat{d}_{CQA}$ was applied to the described dataset of wind time series by considering $C=2$, since the series are associated with two different seasons (winter and summer). The remaining hyperparameters (see Remark \ref{remhyperparameters}) were set as $\mathcal{T}=\{0.1, 0.5, 0.9\}$, $\mathcal{L}=\{1\}$, $r=0.7$ and $m=1.9$. Concerning the three last hyperparameters, the considered values for the selection process were $\mathcal{L}\in\big\{\{1\}, \{1, 2\}, \{1, 2, \ldots, 10\}\big\}$, $m \in \{1.1, 1.2, \ldots, 2\}$, and $r \in \{0.1, 0.2, \ldots, 2\}$.
	
\begin{figure}
		\centering
		\includegraphics[width=0.75\textwidth]{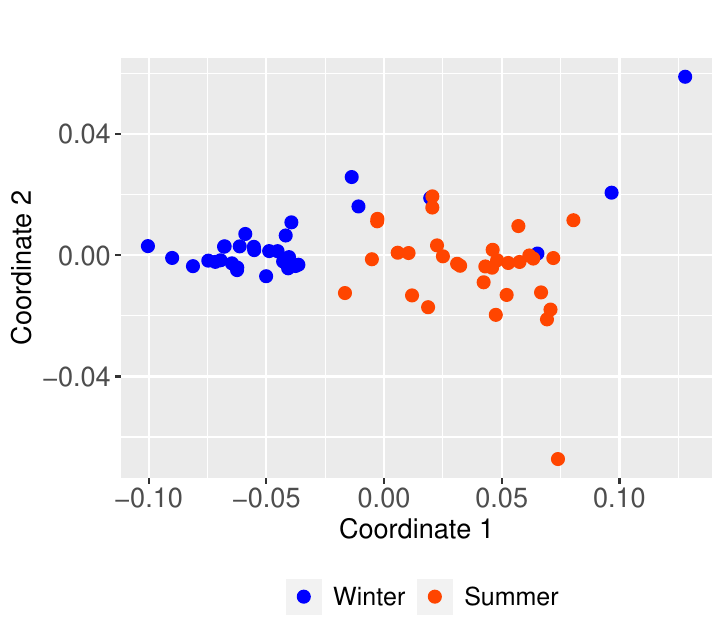}
		\caption{Two-dimensional scaling plane based on distance $\widehat{d}_{CQA}$ for the 64 time series of wind direction in the city of Abha.}
		\label{2ds1}
	\end{figure}

	As a exploratory exercise, we show a 2DS based on the distance $\widehat{d}_{CQA}$, whose computation was carried out using the above values for $\mathcal{T}$, $\mathcal{L}$ and $r$. That way, a projection of the wind time series on a two-dimensional plane preserving the original distances as well as possible is available. The location of the 64 time series in the transformed space is displayed in Figure \ref{2ds1}, where the points were colored according to the season associated with each CTS. The 2DS graph indicates that the distance $\widehat{d}_{CQA}$ is able to detect a clear structure in the dataset according to the underlying season. However, while the winter time series form a rather compact group, the summer time series are more spread out. In both cases, there are some series showing a dynamic behavior which is substantially different from the one characterizing their corresponding seasons. In sum, the configuration in Figure \ref{2ds1} suggests that the fuzzy approach could be appropriate to deal with this dataset.

	Table~\ref{tablefuzzy1} contains the membership degrees produced by the fuzzy $C$-medoids clustering algorithm based on $\widehat{d}_{CQA}$. Superscripts 1 and 2 in the first column indicate the medoid countries according to clusters $\mathcal{C}_1$ and $\mathcal{C}_2$, respectively. For the sake of interpretation, the maximum membership degree of each time series was highlighted in bold as long as it is greater than 0.7.

	\begin{table}
		\centering
		\resizebox{12cm}{!}{\begin{tabular}{lcc|lcc|lcc|lcc} \hline 
				Series	& $\mathcal{C}_1$ & $\mathcal{C}_2$ & Series & $\mathcal{C}_1$ & $\mathcal{C}_2$ & Series & $\mathcal{C}_1$ & $\mathcal{C}_2$ & Series & $\mathcal{C}_1$ & $\mathcal{C}_2$ \\ \hline 
				\textcolor{blue}{Jan 10}	&  \textbf{0.99} & 0.01  &  \textcolor{blue}{Jan 12} &  \textbf{0.99} & 0.01  & \textcolor{blue}{Jan 14} & \textbf{0.99} & 0.01 & \textcolor{blue}{Jan 16} & \textbf{0.99} & 0.01  \\
				\textcolor{blue}{Feb 10} &  \textbf{0.82} & 0.18  & \textcolor{blue}{Feb 12}  &  \textbf{0.99} & 0.01 &  \textcolor{blue}{Feb 14} &  \textbf{0.99} & 0.01& \textcolor{blue}{Feb 16} & \textbf{0.96} & 0.04 \\
				\textcolor{blue}{Mar 10}  &  0.22 & \textbf{0.78}  & \textcolor{blue}{Mar 12} &  \textbf{0.99} & 0.01 & \textcolor{blue}{Mar 14} &  \textbf{0.99} & 0.01 & \textcolor{blue}{Mar 16} & \textbf{0.99} & 0.01 \\
				\textcolor{red}{Jun 10}$^{2}$  &  0.00  & \textbf{1.00}   & \textcolor{red}{Jun 12} &  0.02 & \textbf{0.98} & \textcolor{red}{Jun 14} &  0.01 & \textbf{0.99} & \textcolor{red}{Jun 16} & 0.06 & \textbf{0.94} \\
				\textcolor{red}{Jul 10} &    0.04 & \textbf{0.96}   & \textcolor{red}{Jul 12} &  0.14 & \textbf{0.86} &  \textcolor{red}{Jul 14} &  0.04 & \textbf{0.96}  & \textcolor{red}{Jul 16}  & 0.10 & \textbf{0.90}\\
				\textcolor{red}{Aug 10} &   0.19 & \textbf{0.81}  & \textcolor{red}{Aug 12} &   0.01 & \textbf{0.99} &  \textcolor{red}{Aug 14} &  0.03 & \textbf{0.97} &  \textcolor{red}{Aug 16} & 0.69 & 0.31\\
				\textcolor{red}{Sep 10}  &   0.23 & \textbf{0.77}   & \textcolor{red}{Sep 12}  &   0.04 & \textbf{0.96} &  \textcolor{red}{Sep 14} &  0.07 & \textbf{0.93} & \textcolor{red}{Sep 16} & 0.03 & \textbf{0.97}\\
				\textcolor{blue}{Dec 10} &    \textbf{0.99} & 0.01   &  \textcolor{blue}{Dec 12} &  \textbf{0.99} & 0.01 & \textcolor{blue}{Dec 14} &  0.12 & \textbf{0.88} & \textcolor{blue}{Dec 16} & \textbf{0.99} & 0.01 \\
				\textcolor{blue}{Jan 11} &   \textbf{0.98} & 0.02   &  \textcolor{blue}{Jan 13} &  \textbf{0.99} & 0.01 & \textcolor{blue}{Jan 15 } &  \textbf{0.98} & 0.02& \textcolor{blue}{Jan 17}$^{1}$ & \textbf{1.00}& 0.00 \\
				\textcolor{blue}{Feb 11} &   \textbf{0.99} & 0.01   &  \textcolor{blue}{Feb 13} &  \textbf{0.98} & 0.02 &  \textcolor{blue}{Feb 15} &  \textbf{0.99} & 0.01 &  \textcolor{blue}{Feb 17} & \textbf{0.99} & 0.01\\
				\textcolor{blue}{Mar 11} &   \textbf{0.99} & 0.01    &  \textcolor{blue}{Mar 13} &  \textbf{0.99} & 0.01 & \textcolor{blue}{Mar 15}  &  \textbf{0.81 }& 0.19 & \textcolor{blue}{Mar 17} & \textbf{0.99} & 0.01 \\
				\textcolor{red}{Jun 11}  &   0.03 & \textbf{0.97}   & \textcolor{red}{Jun 13} &  0.06 & \textbf{0.94} & \textcolor{red}{Jun 15} &  0.03 & \textbf{0.97} &  \textcolor{red}{Jun 17} & 0.01 & \textbf{0.99} \\
				\textcolor{red}{Jul 11} &   0.19 & \textbf{0.81}   &   \textcolor{red}{Jul 13} &  \textbf{0.92} & 0.08 &  \textcolor{red}{Jul 15} &  0.02 & \textbf{0.98} & \textcolor{red}{Jul 17} & 0.02 & \textbf{0.98} \\
				\textcolor{red}{Aug 11} &   0.03 & \textbf{0.97 }  &  \textcolor{red}{Aug 13} &  0.02 & \textbf{0.98} &  \textcolor{red}{Aug 15} &  0.01 & \textbf{0.99} & \textcolor{red}{Aug 17} & 0.62 & 0.38 \\
				\textcolor{red}{Sep 11} &   0.08 & \textbf{0.92}   &  \textcolor{red}{Sep 13} &  0.34 & 0.66 & \textcolor{red}{Sep 15} &  0.23 & \textbf{0.77} & \textcolor{red}{Sep 17} & 0.65 & 0.35\\
				\textcolor{blue}{Dec 11} &   0.20 & \textbf{0.80}   &  \textcolor{blue}{Dec 13} &  \textbf{0.99} & 0.01& \textcolor{blue}{Dec 15} &  \textbf{0.97} & 0.03 & \textcolor{blue}{Dec 17} & 0.03 & \textbf{0.97}  \\ \hline
		\end{tabular}}
		\caption{Membership degrees of the 64 wind time series in the city of Abha produced by the fuzzy $C$-medoids clustering algorithm based on the metric $\widehat{d}_{CQA}$ for a 2-cluster partition. The superscripts 1 and 2 are used to indicate the medoid countries for clusters $\mathcal{C}_1$ and $\mathcal{C}_2$, respectively. For each series, the maximum membership degree was highlighted in bold as long as it is greater than 0.7.}
		\label{tablefuzzy1}
	\end{table}
	
	The clustering partition in Table~\ref{tablefuzzy1} is consistent with the corresponding 2DS plot in Figure~\ref{2ds1}. Cluster $\mathcal{C}_1$ contains most of the time series recorded in winter months with high membership degrees, while the opposite happens with cluster $\mathcal{C}_2$ for the time series recorded in summer months. In fact, most of the series exhibit a maximum membership degree close to one, thus indicating that the underlying clustering structure is formed by rather well-separated groups. However, there are four time series displaying a quite uncertain behavior (i.e., with maximum membership degrees below 0.7), namely, the ones associated with the summer months of September 2013, August 2016, August 2017 and September 2017. These time series share the serial dependence structures characterizing both clusters, and they should be individually analyzed in order to understand the reason of their uncertain behavior. Note that this type of insights can be reached due to the fuzzy nature of the partitions, remaining obscured with crisp partitions. In addition, and, as expected from Figure~\ref{2ds1}, there are some winter (summer) time series showing a high membership degree in cluster $\mathcal{C}_2$ ($\mathcal{C}_1$); e.g., March 2010 or July 2013. Such unexpected behavior could indicate the occurrence of uncommon meteorological phenomena during the corresponding months, although extracting such conclusions is beyond the scope of this paper.
	
	For the sake of illustration and comparison purposes, we also obtained the clustering solutions according to the
	fuzzy $C$-medoids model based on the distances $\widehat{d}_{FL}$, $\widehat{d}_{JS}$ and $\widehat{d}_{QA}$ considering $C=2$. The selection of $m$ was carried out in a similar manner than in the above analyses, and the corresponding fuzzy partitions were obtained for the selected values for $m$. Assuming that the true partition is given by the underlying seasons (winter and summer), the ARIF and JIF were obtained for the three mentioned metrics as well as $\widehat{d}_{CQA}$. Note that these quantities evaluate to what extent each dissimilarity is capable of separating the time series according to the corresponding seasons. The results are displayed in Table \ref{indexes64}.

	\begin{table}
		\centering
		\begin{tabular}{ccccc} \hline 
			Metric & $\widehat{d}_{FL}$ & $\widehat{d}_{JS}$  & $\widehat{d}_{CQA}$  & $\widehat{d}_{QA}$  \\ \hline 
			ARIF   & -0.005   & 0.017 & \textbf{0.522} & 0.004 \\ 
			JIF   & 0.326   & 0.337 & \textbf{0.611} & 0.332 \\ \hline  
		\end{tabular}
		\caption{ARIF and JIF obtained by the fuzzy $C$-medoids procedures when grouping the 64 wind time series in the city of Abha. The ground truth is given by the corresponding seasons (winter and summer). The best results are shown in bold.}
		\label{indexes64}
	\end{table}

	According to Table \ref{indexes64}, the proposed metric achieves a quite high value for both clustering indexes, which was expected in view of the fuzzy partition in Table \ref{tablefuzzy1}. In fact, $\widehat{d}_{CQA}$ was the only dissimilarity able to identify the underlying clustering structure to some extent, since the remaining metrics achieve values close to zero for the ARIF, which are indicative of noninformative clustering solutions. The results of this case study justify the need for the metric $\widehat{d}_{CQA}$, since they indicate that: (i) the wind time series exhibit complex forms of dependence that can not be detected through features measuring circular autocorrelation, and (ii) the use of metrics for real-valued time series can lead to erroneous conclusions when the practitioner deals with CTS databases. 
	
\begin{figure}
		\centering
		\includegraphics[width=1\textwidth]{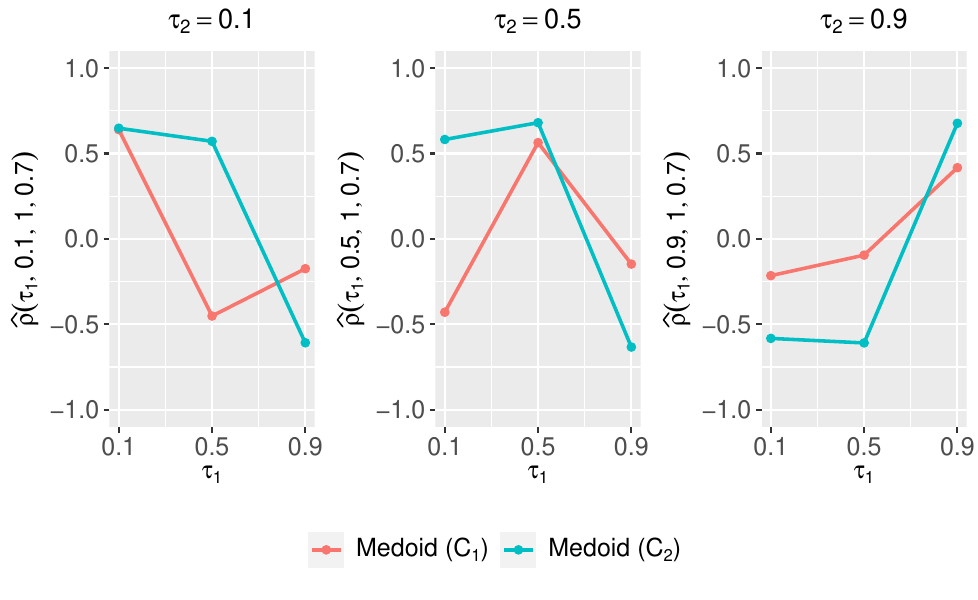}
		\caption{CQA-based estimates for the medoid time series in the first case study for $l=1$ and $r=0.7$.}
		\label{medoids}
	\end{figure}

	When performing partitional clustering of time series, it is common to summarize the characteristics of the elements in each group by means of the corresponding prototypes, namely, the medoids. In our context, the dependence structures characterizing both groups can be represented through the set of features $\big\{\widehat{\rho}(\tau_1, \tau_2, 1, 0.7): \tau_1, \tau_2 =\{0.1, 0.5, 0.9\}\big\}$ associated with each medoid. A graph of these sets is shown in Figure \ref{medoids}, where a different color was used for each medoid. Both medoids show clearly different dependence patterns, with the differences being substantial for features $\widehat{\rho}(0.5, 0.1, 1, 0.7)$ and $\widehat{\rho}(0.1, 0.5, 1, 0.7)$. Although the interpretation of these quantities is not straightforward, a careful analysis of the corresponding values could give the practitioner a general picture on the different dynamics displayed by the wind in both seasons. 
	
	\subsection{Fuzzy clustering of wind direction in different locations}\label{subsectionapplication2}
	
	This second application considers an extended version of the database used in Section \ref{subsectionapplication1}. Specifically, the number of series in that dataset was increased by: (i) considering all the months in the period 2010-2017, instead of only the summer and winter months, and (ii) incorporating the time series associated with one additional city, namely, the city of Makkah, located in the western part of Saudi Arabia. Thus, the new database contains 96 time series associated with each city (Abha and Makkah), for a total of 192 time series. Although the results in Section \ref{subsectionapplication1} indicated that the proposed algorithm can accurately differentiate between winter and summer time series, the main goal of this new case study is to find out whether the procedure can also identify geographical differences in the wind time series, since it is reasonable to think that the dynamic behavior of the wind is dependent on the particular location where it is recorded. Analogous analyses to the ones described in Section \ref{subsectionapplication1} were carried out in the new dataset. As the considered time series were recorded in two different locations, the number of clusters in the fuzzy $C$-medoids algorithm based on $\widehat{d}_{CQA}$ was set to $C=2$.  Concerning the hyperparameters $\mathcal{T}$, $\mathcal{L}$, $m$, and $r$, the values $\mathcal{T}=\{0.1, 0.5, 0.9\}$, $\mathcal{L}=\{1\}$, $r=1$ and $m=2$ were chosen.
	
\begin{figure}
		\centering
		\includegraphics[width=0.75\textwidth]{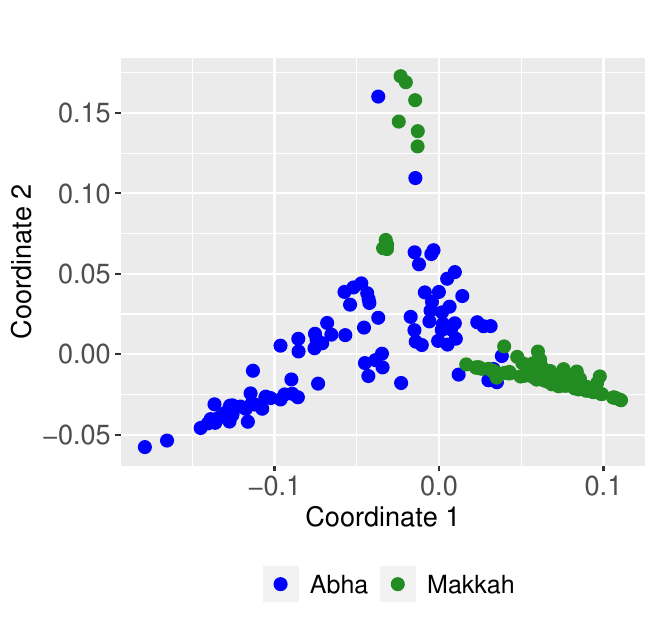}
		\caption{Two-dimensional scaling plane based on distance $\widehat{d}_{CQA}$ for the 192 time series of wind direction in the cities of Abha and Makkah.}
		\label{2ds2}
	\end{figure}

	The 2DS plot based on the distance $\widehat{d}_{CQA}$  for the new dataset is shown in Figure \ref{2ds2}, where the points were colored according to the cities were the time series were recorded. The structure observed in Figure \ref{2ds2} with respect to the ground truth is slightly more ambiguous than the one in the 2DS plot in Figure \ref{2ds1}, probably because no information about the seasons was considered in the construction of the graph. However, clear differences can be seen according to the underlying cities, suggesting that the wind behavior differs among the geographical locations. In fact, most of the time series recorded in Makkah constitute a compact group located in the right part of the graph, while most series associated with Abha are well-separated from this cluster, but showing a higher level of dispersion. Note that, according to Figure \ref{2ds2}, the fuzzy approach can be clearly useful in this application. For instance, some of the time series associated with blue points close to the Makkah cluster are expected to show a high membership degree in this group. In addition, the isolated points located in the top part of the graph could represent time series with an outlying dynamic behavior. In both cases, a careful analysis of the corresponding membership degrees could lead the practitioner to extract meaningful insights from the considered database.

	The values of the clustering indexes associated with $\widehat{d}_{CQA}$ and the alternative metrics in this new case study are shown in Table \ref{indexes192}. Dissimilarities $\widehat{d}_{FL}$ and $\widehat{d}_{JS}$ attain poor results, thus suggesting that circular autocorrelations are again useless to detect the true geographical structure based on the wind time series. On the contrary, dissimilarity $\widehat{d}_{QA}$ achieves the best results, but the corresponding indexes are not substantially different than the ones reached by $\widehat{d}_{CQA}$. In fact, the corresponding ARIF values indicate that both metrics identify both latent groups to some extent. Thus, even though the circular character of the series is not essential here to discriminate between both locations, the proposed distance is able to get similar results than its noncircular counterpart.   
	
	\begin{table}
		\centering
		\begin{tabular}{ccccc} \hline 
			Metric & $\widehat{d}_{FL}$ & $\widehat{d}_{JS}$  & $\widehat{d}_{CQA}$  & $\widehat{d}_{QA}$  \\ \hline 
			ARIF   & 0.104   & 0.012 & 0.419 & \textbf{0.463} \\ 
			JIF   & 0.410   & 0.352 & 0.554 & \textbf{0.582} \\ \hline  
		\end{tabular}
		\caption{ARIF and JIF obtained by the fuzzy $C$-medoids procedures when grouping the 192 wind time series in the cities of Abha and Makkah. The ground truth is given by the corresponding cities. The best results are shown in bold.}
		\label{indexes192}
	\end{table} 
	
	\section{Concluding remarks and future work}\label{sectionconclusions}

	In this work, we have introduced a distance between CTS which automatically takes advantage of the directional character of the time series. Specifically, the metric relies on a new measure of serial dependence considering circular arcs of a given center and radius, which can be seen as an extension of the classical quantile autocorrelation function to the circular setting. We call this measure the circular quantile autocorrelation. A clustering algorithm for circular series was constructed by using the proposed distance in combination with the classical fuzzy $C$-medoids algorithm, which allows for the assignment of gradual memberships of the CTS to the different groups. This is particularly useful when dealing with time series datasets, where different amounts of dissimilarity between the underlying processes or changes in the dynamic behaviors over time are frequent. The algorithm involves some hyperparameters whose selection can be properly carried out by means of some criteria based on hypothesis testing and internal clustering quality indexes. 
	
	To evaluate the performance of the proposed procedure, several simulations were carried out involving scenarios formed by CTS belonging to well-separated clusters and scenarios including a certain amount of uncertainty. Different classes of circular processes were considered. The algorithm was compared with several procedures based on alternative dissimilarities. Specifically, two metrics based on circular autocorrelations and one metric based on the quantile autocorrelation function were considered. The proposed clustering technique showed competitive results with series constructed from classical ARMA processes, and substantially outperformed the alternative procedures when more complex forms of dependence were included in the experiments. 
	
	The usefulness of the proposed algorithm was illustrated by means of two applications involving a dataset of hourly time series of wind direction in several locations of Saudi Arabia between the years 2010 and 2017. In the first case study, we focused on a particular location, split the whole time series into monthly periods, and showed that the procedure is able to differentiate between winter and summer time series, with some time series showing a considerable fuzzy behavior. In the second case study, we considered time series in several locations, and showed that the technique is capable of detecting the corresponding spatial differences to some extent. In sum, both applications indicated that: (i) the new measure of serial dependence can provide a significant understanding about the nature of the time series under study, and (ii) the clustering algorithm can produce a meaningful partition whose membership degrees can provide insights to practitioners about certain time series in the dataset. 
	
	There are some interesting ways through which this work can be extended. First, robust versions of the proposed technique can be constructed by considering the so-called metric, noise, and trimmed approaches \cite{lafuente2020robust, d2016garch, lopez2022quantile2}, which adapt the objective function of the algorithm in a proper manner so that outliers do not negatively affect the resulting partition. Second, a spatial penalization term could be incorporated in the objective function of the procedure in order to deal with CTS data sets containing geographical information \cite{lopez2022spatial, coppi2010fuzzy}, like the one considered in Section \ref{subsectionapplication2}. Third, a spectral counterpart of the circular quantile autocorrelation giving frequency information about the CTS could be defined. Such a measure could lead to the construction of frequency domain clustering algorithms \cite{lopez2021quantile, maharaj2011fuzzy}. Finally, the asymptotic properties of the circular quantile autocorrelation could be derived under certain hypotheses. It would be interesting to address these and further topics in future research.

	\bibliography{mybibfile}       
	

\end{document}